\documentclass[prd,aps,showpacs,,nofootinbib,superscriptaddress,preprintnumbers,epsf,psf]{revtex4}
\usepackage[english]{babel}
\usepackage{pstricks}
\usepackage[dvips]{graphicx}
\usepackage{epsf}
\usepackage{amsmath}
\usepackage{amssymb}
\usepackage{xcolor}

\def\lsim{\raise0.3ex\hbox{$<$\kern-0.75em\raise-1.1ex\hbox{$\sim$}}}
\def\gsim{\raise0.3ex\hbox{$>$\kern-0.75em\raise-1.1ex\hbox{$\sim$}}}

\def\bei{\begin{itemize}}
\def\ei{\end{itemize}}
\def\beqa{\begin{eqnarray}}
\def\eqa{\end{eqnarray}}
\def\bea{\begin{eqnarray}}
\def\eea{\end{eqnarray}}
\def\beas{\begin{eqnarray*}}
\def\eeas{\end{eqnarray*}}
\def\beqas{\begin{eqnarray*}}
\def\eqas{\end{eqnarray*}}

\def\beq{\begin{equation}}
\def\be{\begin{equation}}

\def\ee{\end{equation}}
\def\eq{\end{equation}}
\def\eeq{\end{equation}}

\def\beqd{\begin{displaymath}}
\def\eeqd{\end{displaymath}}
\def\eqd{\end{displaymath}}

\def\beeq{\begin{eqnarray}} \def\eeeq{\end{eqnarray}}

\def\bef{\begin{frame}}

\def\slashchar#1{\setbox0=\hbox{$#1$}
   \dimen0=\wd0
   \setbox1=\hbox{/} \dimen1=\wd1
   \ifdim\dimen0>\dimen1
      \rlap{\hbox to \dimen0{\hfil/\hfil}}
      #1
   \else
      \rlap{\hbox to \dimen1{\hfil$#1$\hfil}}
      /
   \fi}

\newcommand{\C}{\mathbb{C}}

\usepackage{graphicx}
\usepackage{dcolumn}
\usepackage{bm}
\usepackage{epsfig}

\newcommand{\dhd}{{\textstyle d}
\lower.03ex\hbox{\kern-0.38em$^{\scriptstyle-}$}\kern-0.05em{}}
\newcommand{\dbar}{{\textstyle \delta}
\lower.03ex\hbox{\kern-0.38em$^{\scriptstyle-}$}\kern-0.05em{}}

\def\kg{\mathbf{k}}
\def\rg{\textbf{r}}
\def\eg{\boldsymbol{\epsilon}}

\begin{document}


\newcommand{\zbar}{\bar{z}}
\newcommand{\odd}{\mathbb{O}}
\newcommand{\pom}{\mathbb{P}}
 \makeatletter
 \def\preprint#1{ \def\@preprint{\noindent\hfill\hbox{#1}\vskip 10pt}}


\noindent\hfill\hbox{\begin{tabular}{l}
      CPHT RR009.0311\\
       LPT 11-26
 \end{tabular}}\vskip 10pt



\title{A phenomenological study of helicity amplitudes of high energy exclusive leptoproduction of the $\rho$ meson}

\author{I.~V.~Anikin}
\affiliation{Bogoliubov Laboratory of Theoretical Physics, JINR,
             141980 Dubna, Russia
and \\
Institute for Theoretical Physics, University of Regensburg,
D-93040 Regensburg, Germany}

\author{A. Besse}
\affiliation{LPT, Universit{\'e} Paris-Sud, CNRS, 91405, Orsay, France}

\author{D.Yu.~Ivanov}
\affiliation{Sobolev Institute of Mathematics and Novosibirsk State University, 630090 Novosibirsk, Russia}

\author{B.~Pire}
\affiliation{CPHT, {\'E}cole Polytechnique, CNRS, 91128 Palaiseau Cedex, France}

\author{L.~Szymanowski}
\affiliation{Soltan Institute for Nuclear Studies, PL-00-681 Warsaw, Poland}

\author{S.~Wallon}
\affiliation{LPT, Universit{\'e} Paris-Sud, CNRS, 91405, Orsay, France and \\
UPMC Univ. Paris 06, facult\'e de physique, 4 place Jussieu, 75252 Paris Cedex 05, France}

\begin{abstract}

We apply a previously developed scheme  to
consistently include the twist-3 distribution
amplitudes for transversely polarized $\rho$ mesons in order
  to evaluate, in the framework of $k_T$ factorization, the helicity amplitudes
for exclusive leptoproduction of a light vector meson,
at leading order in $\alpha_s$.
  We compare our results with
high energy experimental data for the ratios of helicity amplitudes $T_{11} /T_{00}$ and $T_{01} /T_{00}$ and
get a good description of the data.

\end{abstract}
%


\maketitle
\narrowtext


\noindent


\section{Introduction}
\label{sec:intro}

Exclusive leptoproduction of vector mesons has been the subject of significant progress in the last 20 years,
in particular, in the hard regime where a highly virtual photon exchange allows one to separate a short distance
dominated amplitude of a hard subprocess from suitably defined hadronic objects. Experimental knowledge
 has
been gathered in a wide range of center-of-mass energies, from a few GeV at JLab to hundreds of GeV at
the HERA collider.
Following the pioneering NMC \cite{Arneodo:1994id} and E665
\cite{Adams:1997bh} experiments, the HERA collaborations H1 and ZEUS measured $\rho$-meson electroproduction  \cite{Breitweg:1998nh,  Breitweg:1999fm}. COMPASS also measured the same reaction in an intermediate energy range
\cite{Alexakhin:2007mw}. Lower energy data
have been extracted at HERMES  \cite{Airapetian:2009ae, Borissov:2009zz} and JLab \cite{Morrow:2008ek}.

The H1 and ZEUS collaborations have recently provided a complete analysis  \cite{Chekanov:2007zr, Aaron:2009xp} of spin density matrix elements describing the hard exclusive productions of the $\rho$ and the $\phi$ vector mesons $V$ in the process
\begin{equation}
\label{process}
\gamma^{*}(\lambda_{\gamma}) \, p \rightarrow V(\lambda_{V}) \, p \,,
\end{equation}
which can be expressed in terms of  helicity amplitudes $T_{\lambda_{V}\lambda_{\gamma}}$ ($\lambda_{\gamma}$, $\lambda_{V}$ : polarizations of the virtual photon and the vector meson).

The ZEUS collaboration \cite{Chekanov:2007zr} has provided data for different photon virtualities $-Q^2$, i.e. for
 $2 < Q^2 < 160$ GeV$^2, \, 32 < W < 180$ GeV $(\;|t| < 1$ GeV$^2),$ while the H1 collaboration \cite{Aaron:2009xp} has analyzed data
in the range
$2.5 < Q^2 < 60$ GeV$^2, \, 35 <  W < 180$ GeV  $\;(|t| < 3$ GeV$^2)\,,$
where $W$ is the center-of-mass energy of the  virtual photon-proton system.

The main features of the HERA data are as follows: specific $Q^2$ scaling, and $t$ and $W$ dependence of the cross sections
(features distinct from those in soft diffractive reactions) strongly support the idea that the dominant mechanism of the diffractive process (\ref{process}) is the scattering of a small transverse-size, $\sim 1/Q$,
colorless dipole on the proton target. This justifies the use of perturbative QCD methods for the description of the process (\ref{process}).

On the theoretical side, three main approaches have been developed.  The first two, a $k_T$-factorization approach and a dipole approach,  are applicable at high energy, $W\gg Q \gg \Lambda_{QCD}$. They are both related to a
Regge inspired $k_T$-factorization scheme \cite{Cheng:1970ef, FL, GFL, Catani:1990xk, Catani:1990eg, Collins:1991ty, Levin:1991ry}, which basically writes the scattering amplitude in terms of two impact
factors one, in our case,  for the $\gamma^* - \rho$ transition and the other one for the nucleon to nucleon transition,
with, at leading order, a two "Reggeized" gluon exchange in the $t$-channel. The Balitsky-Fadin-Kuraev-Lipatov (BFKL)  evolution, known at leading order (LLx) \cite{Fadin:1975cb, Kuraev:1976ge, Kuraev:1977fs, Balitsky:1978ic} and next-to-leading (NLLx) order \cite{Fadin:1996tb, Camici:1997ij, Ciafaloni:1998gs, Fadin:1998py}, can then be applied
to account for a specific large energy QCD resummation.
The dipole approach is  based on  the formulation of similar ideas, not in $k_T$ but in transverse coordinate space \cite{Mueller:1989st, Nikolaev:1990ja}; this scheme is especially suitable to account for nonlinear evolution and gluon saturation effects.
The third approach, valid also for $W\sim Q$, was initiated in \cite{Brodsky:1994kf} and \cite{Frankfurt:1995jw}.
It is based on
the collinear QCD factorization scheme \cite{Collins:1996fb, Radyushkin:1997ki}; the amplitude is given as  a convolution of quark or gluon generalized parton
distributions (GPDs) in the nucleon, the $\rho$-meson distribution amplitude (DA), and a perturbatively calculable
hard scattering amplitude. GPD evolution equations resum the collinear gluon effects. The DAs are subject to specific QCD evolution equations \cite{Farrar:1979aw,  Lepage:1979zb, Efremov:1979qk}.

Though the collinear factorization approach allows us to calculate perturbative corrections to the leading twist longitudinal amplitude (see \cite{Ivanov:2004zv} for NLO), when dealing with transversely polarized vector mesons, one faces end-point singularity problems. Consequently, this does not allow us to study polarization effects in diffractive $\rho$-meson electroproduction in a model-independent way within the collinear factorization approach. An improved collinear approximation scheme has been proposed based
on Sudakov  factors \cite{Li:1992nu}, which allows us to overcome end-point singularity problems, and has been applied to $\rho$-electroproduction \cite{Vanderhaeghen:1999xj, Goloskokov:2005sd, Goloskokov:2006hr, Goloskokov:2007nt}.

In this study, we consider polarization effects for reaction (\ref{process}) in the high energy region, $s=W^2\gg Q^2\gg \Lambda^2_{QCD}$,
working within the $k_T$-factorization approach, where one can represent the forward helicity amplitudes as\footnote{We use boldface letters for Euclidean two-dimensional transverse vectors.}
\begin{equation}
T_{\lambda_{\rho}\lambda_{\gamma}}(s;Q^2) \propto is\int \frac{d^2\textbf{k}}{(\textbf{k}^2)^2} \Phi^{\gamma^{*}(\lambda_{\gamma}) \rightarrow \rho(\lambda_{\rho})}(\textbf{k}^2,Q^2)\, {\cal F}(x,\textbf{k}^2 )\,,
\quad x=\frac{Q^2}{s} \, ,
\label{kt-fact}
\end{equation}
where $\Phi^{\gamma^{*}(\lambda_{\gamma}) \rightarrow \rho(\lambda_{\rho})}(\textbf{k}^2,Q^2)$ is an impact factor describing the virtual photon to $\rho$-meson transition, and ${\cal F}(x,\textbf{k}^2 )$ is an unintegrated gluon density, which at the Born order is simply related to the
proton-proton impact factor, as described below. Here, $\textbf{k}$ is the transverse momentum of the $t-$channel exchanged gluons.

The impact factor $\Phi^{\gamma^{*}(\lambda_{\gamma}) \rightarrow \rho(\lambda_{\rho})}(\textbf{k}^2,Q^2)$ vanishes at $\textbf{k}\to 0,$
which guarantees the convergence of the integral in Eq.(\ref{kt-fact}) on the lower limit\footnote{This property of the impact factors is universal and related to the gauge invariance \cite{Frolov:1970ij,Fadin:1999qc}.
It is a consequence of QCD gauge invariance and is in accordance with the
general statement of the Kinoshita, Lee and Nauenberg theorem which guarantees the  infrared finiteness
of  amplitudes in the case of the scattering of colorless objects.}. In fact, $\Phi_{\gamma^*_L \rightarrow \rho_L}\sim \textbf{k}^2/Q^2$ at $|\textbf{k}|\ll Q$, which allows us to express the longitudinal amplitude in the collinear limit in terms of the usual gluonic parton distribution function, 
$$x\, g(x,Q^2)=\int\limits^{Q^2} \frac{d\textbf{k}^2}{\textbf{k}^2} {\cal F}(x,\textbf{k}) \, .$$
In the case of the transverse amplitude, the situation in the collinear limit is different due to the above-mentioned  end-point
singularities.  At $|\textbf{k}|\ll Q$ the  transverse impact factor  $\Phi_{\gamma^*_T \rightarrow \rho_T}\sim (\textbf{k}^2/Q^2) \ln \left(Q^2/\textbf{k}^2\right)$ and the transverse amplitude cannot be expressed in terms of the
gluonic parton distribution function.  Nevertheless, both longitudinal and transverse amplitudes can be calculated within 
the $k_T$-factorization description (\ref{kt-fact}). For the transverse amplitude the end-point singularities are naturally regularized by the transverse
momenta of the $t-$channel gluons \cite{Ivanov:1998gk, Anikin:2009hk, Anikin:2009bf}.

Moreover, at large photon virtuality providing the hard scale, the $\gamma^* - \rho$ impact factors can be calculated 
in a model-independent way using QCD twist expansion in the region $\textbf{k}^2\gg \Lambda_{QCD}^2$. Such a calculation  involves the
$\rho$-meson DAs as nonperturbative inputs. The principle point here is that the region $\textbf{k}^2\gg \Lambda_{QCD}^2$ gives the dominant contribution to the amplitudes in the integral (\ref{kt-fact}). Below we introduce an
explicit cutoff for transverse momenta of $t-$channel gluons to clarify this point.
The calculation of the impact factors for $\Phi_{\gamma^{*}_L \rightarrow \rho_L}$, $\Phi_{\gamma^{*}_T \rightarrow \rho_L}$
is standard at the twist-2 level \cite{Ginzburg:1985tp}, while $\Phi_{\gamma^{*}_T \rightarrow \rho_T}$ was only recently computed \cite{Anikin:2009hk,Anikin:2009bf} (for the forward case $t = t_{min}$),  up to twist-3, including two- and  three-body correlators, which contribute here on an equal footing.

In our study we use results \cite{Ginzburg:1985tp, Anikin:2009bf} for the $\Phi_{\gamma^*_L \rightarrow \rho_L}$ and  $\Phi_{\gamma^*_T \rightarrow \rho_T}$ impact factors and a phenomenological model \cite{Gunion:1976iy} for the proton-proton impact factor. This model involves a single energy scale parameter $M$, and is equivalent in our calculation to a specific assumption for the  $\textbf{k}$ dependence shape of the unintegrated gluon distribution ${\cal F}(x,\textbf{k})$.

Our approach is close in spirit to the calculations performed in Refs.~\cite{Nemchik:1996cw, Forshaw:2003ki, Kowalski:2006hc, Forshaw:2010py, Forshaw:2011yj} within the dipole approach, where 
amplitudes are related to the light-cone wave functions $\phi(z,\textbf{r})$.
Collinear DAs are integrals of  light-cone wave functions in momentum space over the relative  transverse momentum
conjugated to $\textbf{r}.$
 The light-cone wave functions are complicated objects, and in practice, 
 their dependence on the longitudinal momentum fraction $z$ and transverse coordinate separation $\textbf{r}$ variables, which describe the $q\bar q$ dipole, should be modeled. 
Our main point here is that one can assume  the dominant physical mechanism for production of both longitudinal and transversely polarized mesons to be the scattering of small transverse-size quark-antiquark and
quark-antiquark-gluon colorless states on the target. This allows one to calculate corresponding helicity amplitudes 
in a model-independent way, using the natural light-cone QCD language --  twist-2 and twist-3 DAs.




This paper is organized as follows. In Sec.~\ref{sec:IF}, we recall some results of Refs.~\cite{Anikin:2009bf} and \cite{Ball:1998sk} about the impact factors and distribution amplitudes for the $\rho^{0}$ meson, which we need for computing helicity amplitudes. 
In Sec.~\ref{sec:helicity_ampl}, we
describe first the model for the proton-proton impact factor. Then we
compare with HERA data the ratios of helicity amplitudes $T_{11}/T_{00}$,
calculated both in the Wandzura-Wilczek (WW) approximation and with the account of
the genuine contribution. We also compare our predictions for the
$T_{01}/T_{00}$ ratio with the data of HERA.
 We obtain a good description of these two ratios. For both observables, we discuss  the effect of the energy scale $M^2$  for the  proton-proton impact factor, as well as
the sensitivity to the infrared region of $t$-channel gluon momenta.



\section{Impact factors $\gamma^{*}(\lambda_{\gamma}) \rightarrow \rho(\lambda_{\rho})$}
\label{sec:IF}

\subsection{Impact factor representation}
\label{subsec:IF_rep}

\begin{figure}[h]
\centerline{\raisebox{0cm}{\epsfig{file=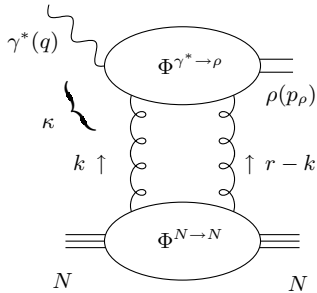,width=4.2cm,clip=}}}
\caption{Impact factor representation of the $\gamma^* \, N  \to  \rho \, N$ scattering amplitude.}
\label{Fig:impact_fact}
\end{figure}

In the impact factor representation at the Born order, the amplitude of the exclusive process $\gamma^{*}(\lambda_{\gamma}) \, N \rightarrow \rho (\lambda_{\rho}) \, N$ reads
\begin{equation}
\label{defImpactRep}
 T_{\lambda_{\rho}\lambda_{\gamma}}(\textbf{r};Q , M) = is\int \frac{d^2\textbf{k}}{(2\pi)^2}\frac{1}{\textbf{k}^2(\textbf{k}-\textbf{r})^2} \Phi^{N \rightarrow N} (\textbf{k},\textbf{r};M^2)\Phi^{\gamma^{*}(\lambda_{\gamma}) \rightarrow \rho(\lambda_{\rho})}(\textbf{k},\textbf{r};Q^2)\,,
 \end{equation}
as illustrated in Fig.~\ref{Fig:impact_fact}.
Note that in this representation,  no skewness
effect is taken into account.
The $\gamma^{*}(\lambda_{\gamma}) \rightarrow \rho(\lambda_{\rho})$ impact factor
$ \Phi^{\gamma^{*}(\lambda_{\gamma}) \rightarrow \rho(\lambda_{\rho})}$ is defined through the discontinuity of the  $S$ matrix element for $\gamma^{*}(\lambda_{\gamma};q) g(k) \rightarrow g(r-k) \rho(\lambda_{\rho};p_{\rho})$
as
\beq
\label{ImpactDisc}
\Phi^{\gamma^{*}(\lambda_{\gamma}) \rightarrow \rho(\lambda_{\rho})}=\frac{1}{2 s}\int \frac{d \kappa}{2 \pi} \, {\rm Disc}_\kappa  \left( S_{\mu \nu}^{\gamma^* g \to \rho g} \, p_2^\mu \, p_2^\nu \frac{2}{s} \right)\,,
\eq
where $
\kappa =(k+q)^2$\,. In Eqs.(\ref{defImpactRep}) and (\ref{ImpactDisc}) the momenta $q$ and $p_\rho$ are parametrized via Sudakov decompositions in terms of two
 lightlike vectors $p_1$ and $p_2$ such that $2\, p_1.p_2=s$, as
\begin{equation}
q=p_1-\frac{Q^2}{s}p_2 \quad {\rm and } \quad p_{\rho}=p_1+\frac{m_{\rho}^2-t+t_{min}}{s} p_2 + r_\perp\,,
\end{equation}
where $Q^2=-q^2 >> \Lambda_{QCD}^2$ is the virtuality of the photon, which justifies the use of perturbation theory, and $m_{\rho}$ is the mass of the $\rho$ meson. The impact
$ \Phi^{N \rightarrow N}$ in Eq.(\ref{defImpactRep}) cannot be computed within perturbation theory, and we will use a model described in Sec.\ref{subsec:Model}.
Note that due to QCD gauge invariance, both  impact factors should vanish when either one of the transverse momenta of the $t-$channel exchanged gluon goes to zero, $\textbf{k} \to 0$ or $\textbf{r-k} \to 0\,.$

\begin{figure}[h]
\begin{tabular}{cccc}
\includegraphics[width=5.8cm]{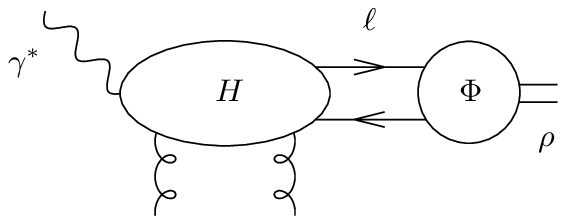}&\hspace{.2cm}
\raisebox{1.2cm}{+}&\hspace{.3cm}\includegraphics[width=5.8cm]{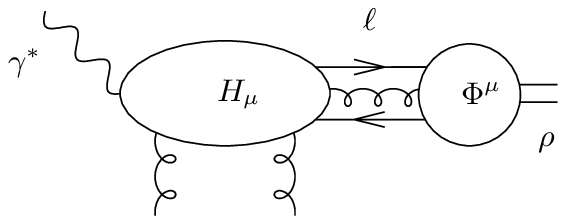}
&\hspace{.2cm}\raisebox{1.2cm}{$+ \cdots$}
\end{tabular}
\caption{Two- and three-parton correlators attached to a hard scattering amplitude in the  specific case of the $\gamma^* \to \rho$ impact factor,
where vertical lines are hard $t-$ channel gluons in the color
singlet state.}
\label{Fig:NonFactorized}
\end{figure}

The computation of the  $\gamma^* \to \rho$ impact factor is performed within collinear factorization of QCD.
The
dominant contribution corresponds to the $\gamma^*_L \to \rho_L$ transition (twist 2), while the other transitions  are power suppressed.
The $\gamma^*_L \to \rho_L$ and  $\gamma^*_T \to \rho_L$ impact factors were computed a long time ago \cite{Ginzburg:1985tp}, while a consistent treatment of the twist-3
$\gamma^*_T \to \rho_T$ impact factor has been performed only recently in Ref.~\cite{Anikin:2009bf}.
It is based on the light-cone collinear factorization (LCCF) beyond the leading twist, applied to the amplitudes $\gamma^{*}(\lambda_{\gamma})g(k) \rightarrow g(r-k)\rho(\lambda_{\rho})$, symbolically illustrated in Fig.~\ref{Fig:NonFactorized}.
Each of these scattering amplitudes is the sum of the convolution of a hard part (denoted by $H$ and $H_\mu$ for two- and three-body contributions, respectively) that corresponds to the transition of the virtual photon into the constituents of the $\rho$ meson and their interactions with off-shell gluons of the $t$ channel, and a soft part (denoted by $\Phi$ and $\Phi^\mu$). As the photon is highly virtual,
this convolution reduces to a factorized form, expressed as a convolution in the longitudinal momentum of  hard parts in collinear kinematics  with DAs.

\subsection{Distribution amplitudes in the LCCF parametrization}
\label{subsec:DA_LCCF}

 Our accuracy is limited to dominant contributions both for $\gamma^*_L \to \rho_L$ (twist-2) and $\gamma^*_T \to \rho_T$ (twist-3) transitions; therefore, only leading terms of the expansion in $1/Q$ in both amplitudes are kept. Hence, only two-body (quark-antiquark) and three-body (quark-antiquark gluon) nonlocal operators are involved.
Correlators and distribution amplitudes depend on a factorization scale $\mu$. We have to take into account this dependence to compare our results with experimental data.

The seven chiral-even \footnote{The chiral-odd twist-2 DA for the transversely polarized $\rho$ meson does
not contribute to the process considered at the accuracy discussed here. This is also true in the approach
based on collinear factorization of generalized parton distributions \cite{Diehl:1998pd, Collins:1999un}.}
$\rho$-meson DAs  up to twist 3 are defined by  matrix elements of nonlocal
light-cone operators.
Let us introduce $p$ and $n$, two light-cone vectors such that $p_{\rho}\approx p$ at twist 3 and $p \cdot n = 1$. The polarization of the out-going $\rho$ meson is denoted by $e^*$.

The two-body correlators are parametrized as\footnote{In the approximation where the mass of the quarks is neglected with respect to the mass of the $\rho$ meson.} \cite{Anikin:2009bf}
\beqa
\label{defDA_V}
 \left\langle \rho(p_{\rho}) \left|\bar{\psi}(z)\gamma_{\mu}\psi(0)\right|0\right\rangle & =& m_{\rho} f_{\rho}  \int^{1}_{0} dy \, e^{i yp.z}  [\varphi_1(y;\mu^2) (e^*.n)p_{\mu}+\varphi_3(y;\mu^2)e_{T\mu}^*] \\
\label{defDA_Vder}
 \left\langle \rho(p_{\rho}) \left|\bar{\psi}(z)\gamma_5\gamma_{\mu}\psi(0)\right|0\right\rangle & =&i \, m_{\rho} f_{\rho} \, \varepsilon_{\mu\alpha\beta\delta} \, e_T^{*\alpha} p^{\beta}n^{\delta} \int^{1}_{0} dy \, e^{i yp.z}  \varphi_A(y;\mu^2)\\
\label{defDA_A}
 \left\langle \rho(p_{\rho}) \left|\bar{\psi}(z)\gamma_{\mu}i
\stackrel{\longleftrightarrow}
{\partial^T_{\alpha}}
\psi(0)\right|0\right\rangle & =&m_{\rho} f_{\rho} \, p_{\mu}e^{*}_{T\alpha} \int^{1}_{0} dy \,  e^{i yp.z}  \varphi_1^T(y;\mu^2)\\
\label{defDA_Ader}
 \left\langle \rho(p_{\rho}) \left|\bar{\psi}(z)\gamma_5\gamma_{\mu}i\stackrel{\longleftrightarrow}
{\partial^T_{\alpha}}
\psi(0)\right|0\right\rangle & =& i \, m_{\rho} f_{\rho} \, p_{\mu}\, \varepsilon_{\alpha\lambda\beta\delta}\,
e_T^{*\lambda} p^{\beta}\,n^{\delta}\int^{1}_{0} dy \, e^{i yp.z}  \varphi_A^T(y;\mu^2)
\,,
 \eqa
where $y$ and $\bar{y}=1-y$ are, respectively, the momentum fractions of the quark and the antiquark,
while the three-body correlators are expanded as
 \beqa
\label{defDA_B}
\hspace{-.6cm}\left\langle \rho(p_{\rho}) \left|\bar{\psi}(z_1)\gamma_{\mu} g A^T_{\alpha}(z_2)\psi(0)\right|0\right\rangle &\!\! \!=&\!\!\!m_{\rho} f^{V}_{3\rho}(\mu^2) \,
p_{\mu} \, e_{T\alpha}^*
 \int^{1}_{0}dy_2\int^{y_2}_{0} d y_1 e^{i y_1p.z_1+i(y_2-y_1)p.z_2}  B(y_1,y_2;\mu^2) \\
\label{defDA_D}
\hspace{-.6cm}\left\langle \rho(p_{\rho}) \left|\bar{\psi}(z_1)\gamma_5\gamma_{\mu} g A^T_{\alpha}(z_2)\psi(0)\right|0\right\rangle &\!\! \!=&\!\!\!i \, m_{\rho} f^{A}_{3\rho}(\mu^2) \,
\varepsilon_{\alpha\lambda\beta\delta}\,
e_T^{*\lambda}p^{\beta}n^{\delta}\!\!
 \int^{1}_{0}\!\!\!\!dy_2 \! \int^{y_2}_{0} \!\!\!\!d y_1 \, e^{i y_1p.z_1+i(y_2-y_1)p.z_2}  D(y_1,y_2;\mu^2) \,,\,\,
\eqa
where
 $y_1$, $\bar{y_2},$ and $y_g = y_2-y_1$ are, respectively, the momentum fractions of the quark, the antiquark, and the gluon.
We used the standard notation
$\stackrel{\longleftrightarrow}
{\partial_{\rho}}=\frac{1}{2}(\stackrel{\longrightarrow}
{\partial_{\rho}}-\stackrel{\longleftarrow}{\partial_{\rho}})\,.$
 Normalizations and symmetry properties of the distribution amplitudes are recalled in Appendix \ref{subsec:DA}, and they are used to simplify helicity amplitudes in the forthcoming computations. For later use,
 we define  the following combinations of three-body distribution amplitudes:
\beqa
\label{defSM}
M(y_1,y_2;\mu^2)&=&\zeta^{V}_{3\rho}(\mu^2) \, B(y_1,y_2;\mu^2)-\zeta^{A}_{3\rho}(\mu^2) \, D(y_1,y_2;\mu^2)\,, \nonumber
\\
S(y_1,y_2;\mu^2)&=&\zeta^{V}_{3\rho}(\mu^2) \, B(y_1,y_2;\mu^2)+\zeta^{A}_{3\rho}(\mu^2) \,D(y_1,y_2;\mu^2)
\eqa
where $\zeta^V_{3\rho}(\mu^2)$ and $\zeta^{A}_{3\rho}(\mu^2)$ are dimensionless coupling constants:
\beq
\label{def_zeta}
\zeta^{V}_{3\rho}(\mu^2)=\frac{f^V_{3\rho}(\mu^2)}{f_{\rho}} \qquad
\zeta^{A}_{3\rho}(\mu^2)=\frac{f^A_{3\rho}(\mu^2)}{f_{\rho}}\,.
 \eq

\subsection{Reduction of DAs to a minimal set $\varphi_1$, $B$, $D$}
\label{subsec:Red}

DAs are not independent; they are linked by linear differential relations derived from equations of motion and $n$ independency \cite{Anikin:2009hk,Anikin:2009bf}. The solutions for $\varphi_P(y)\equiv\{\varphi_3, \varphi_A, \varphi_1^T, \varphi_A^T\}$ are the sum of the solutions in the so-called WW approximation and  of genuine solutions:
\begin{equation}
 \varphi_P(y)=\varphi^{WW}_P(y)+\varphi^{gen}_P(y)\,.
 \end{equation}
The WW approximation consists in neglecting the contribution from three-body operators by taking $B(y_1,y_2;\mu^2)=D(y_1,y_2;\mu^2)=0$. $\varphi_P^{WW}(y)$ are then only functions of $\varphi_1(y)$:
\beqa
\label{phi3_WW}
 \varphi^{WW}_{3}(y;\mu^2)&=&\frac{1}{2} \left[\int^{y}_{0}du \frac{\varphi_1(u;\mu^2)}{\bar{u}}+\int^{1}_{y} du \frac{\varphi_1(u;\mu^2)}{u}\right]\,,\\
\label{phiA_WW}
 \varphi^{WW}_{A}(y;\mu^2)&=&\frac{1}{2}\left[\int^{y}_{0}du \frac{\varphi_1(u;\mu^2)}{\bar{u}}-\int^{1}_{y} du \frac{\varphi_1(u;\mu^2)}{u}\right]\,,\\
\label{phiAT_WW}
\varphi^{WW}_{AT}(y;\mu^2)&=&-\frac{1}{2}\left[\bar{y} \int^{y}_{0}du \frac{\varphi_1(u;\mu^2)}{\bar{u}}+y\int^{1}_{y} du \frac{\varphi_1(u;\mu^2)}{u}\right]\,,\\
\label{phi1T_WW}
\varphi^{WW}_{1T}(y;\mu^2)&=&\frac{1}{2}\left[-\bar{y} \int^{y}_{0}du \frac{\varphi_1(u;\mu^2)}{\bar{u}}+y \int^{1}_{y} du \frac{\varphi_1(u;\mu^2)}{u}\right]\,.
\eqa
Genuine solutions only depend on $\{B(y_1,y_2;\mu^2), D(y_1,y_2;\mu^2)\}$ or the combinations $\{S(y_1,y_2;\mu^2), M(y_1,y_2;\mu^2)\}$:
\beqa
\label{exp_phi3gen}
\varphi^{gen}_3(y;\mu^2)&=&\frac{1}{2}\left[\int^{1}_{\bar{y}}d u\,\frac{A(u;\mu^2)}{u}+\int^{1}_{y}d u \, \frac{A(u;\mu^2)}{u} \right]
\\
\label{exp_phiAgen}
 \varphi^{gen}_A(y;\mu^2)&=&\frac{1}{2}\left[\int^{1}_{\bar{y}}d u \,\frac{A(u;\mu^2)}{u}-\int^{1}_{y}d u \,\frac{A(u;\mu^2)}{u} \right]\,,
 \eqa
where $A(u;\mu^2)$ has the compact form
\begin{equation}
A(u;\mu^2)=\int^{u}_{0}d y_2\left[\frac{1}{y_2-u}-\partial_u\right]M(y_2,u;\mu^2)+\int^{1}_{u}dy_2 \frac{1}{y_2-u}M(u,y_2;\mu^2)
\end{equation}
and  it obeys the conditions
\begin{equation}
\int^{1}_{0}du \, A(u;\mu^2)=0 \quad {\rm and} \quad \int^{1}_{0}du \,\bar{u} \, A(u;\mu^2)=0\,,
\end{equation}
coming, respectively, from the constraints
\beq
\label{int_phi3_phiA}
\int^1_0 \varphi_3^{gen}(y; \, \mu^2)  \, dy =0 \quad {\rm and} \quad  \int^1_0 (y-\bar{y}) \, \varphi_A^{gen}(y; \, \mu^2)  \, dy =0\,.
\eq
Equations (\ref{exp_phi3gen}) and (\ref{exp_phiAgen}) determine the expressions of $\varphi^{gen}_{1T}(y;\mu^2)$ and $\varphi^{gen}_{AT}(y;\mu^2)$ as
\beqa
\label{exp_phi1Tgen}
\varphi^{gen}_{1T}(y;\mu^2)&=&\int^{y}_{0}du\, \varphi^{gen}_3(u;\mu^2)-\frac{1}{2} \int^{y}_{0}dy_1\int^{1}_{y}dy_2 \frac{S(y_1,y_2;\mu^2)+M(y_1,y_2;\mu^2)}{y_2-y_1}\,,
\\
\label{exp_phiATgen}
\varphi^{gen}_{AT}(y;\mu^2)&=&\int^{y}_{0}du\,\varphi^{gen}_A(u;\mu^2)-\frac{1}{2} \int^{y}_{0}dy_1\int^{1}_{y}dy_2 \frac{S(y_1,y_2;\mu^2)-M(y_1,y_2;\mu^2)}{y_2-y_1}\,.
\eqa
The correspondence between our set of DAs and the one defined in Ref.~\cite{Ball:1998sk}
is achieved through the following dictionary derived in Ref.~\cite{Anikin:2009bf}. It reads,
for the two-body vector DAs,
\begin{eqnarray}
\label{relBBvector}
&&\varphi_1(y)=
\phi_{\parallel}(y) ,
\quad
\varphi_3(y)=
 g_\perp^{(v)}(y) \,,
\end{eqnarray}
and for the axial DA,
\begin{eqnarray}
\label{relBBaxial}
&&\varphi_A(y) =
-\frac{1}{4} \, \frac{\partial g_\perp^{(a)}(y)}{\partial y}\,.
\end{eqnarray}
For the three-body DAs, the identification is
\begin{eqnarray}
\label{DictB_D}
 B(y_1,\,y_2)=-\frac{V(y_1, \, 1-y_2)}{y_2-y_1} \quad {\rm and } \quad
D(y_1,\,y_2)=-\frac{A(y_1, \, 1-y_2)}{y_2-y_1}\,.
\end{eqnarray}
Explicit forms for $\varphi_1$, $B$, and $D$ are obtained with the help of the results of Ref.~\cite{Ball:1998sk} obtained within the QCD sum rules approach.
The first terms of the expansion in the momentum fractions of the three independent DAs thus have the form
 \beqa
\label{asymp_phi1}
 \varphi_1(y,\mu^2)&=&6y\bar{y}(1+a_2(\mu^2)\frac{3}{2} (5 (y-\bar{y})^2-1))\,,\\
\label{asymp_B}
 B(y_1,y_2;\mu^2)&=&-5040 y_1 \bar{y}_2 (y_1-\bar{y}_2) (y_2-y_1)\,,\\
D(y_1,y_2;\mu^2)&=&-360 y_1\bar{y}_2(y_2-y_1) (1+\frac{\omega^{A}_{\{1,0\}}(\mu^2)}{2}(7(y_2-y_1)-3))\,.
\label{asymp_D}
\eqa
The dependences on the renormalization scale $\mu^2$ of the coupling constants $a_2$, $\omega^{A}_{\{1,0\}}$, $\zeta^{A}_{3\rho}$, and $\zeta^{V}_{3\rho}$ are given in Ref.~\cite{Ball:1998sk}. 
In Appendix B we present both the evolution equations and the  
values of these constants at $\mu^2=1$~GeV$^2$ used in our analysis,  as well as the dependence on $\mu^2$ of the DAs.



\subsection{Impact factors}
\label{subsec:IF}

In the Sudakov basis,
the longitudinal and
 transverse polarizations   of the photon are\footnote{In Ref.~\cite{Anikin:2009bf} we took  $\epsilon^{\pm}=\mp\frac{i}{\sqrt{2}}\left(0, 1,\pm i, 0\right)$, which we change here for consistency with the usual experimental conventions \cite{:2010dh}.}
\beq
\label{def_polL_polT}
e_{\gamma L}^{\mu}=\frac{1}{Q} (p_1^{\mu}+\frac{Q^2}{s}p_2^{\mu})\,,
\qquad
 \epsilon^{\pm}=\frac{1}{\sqrt{2}}\left(0, \mp1,- i, 0\right)\,.
\eq
For $t=t_{min}$ the same parametrization will be used for the $\rho$-meson polarization with $Q^2 \to -m_{\rho}^2$ and $Q \to m_{\rho}$.
We introduce the notations: $\alpha = \frac{\textbf{k}^2}{Q^2}$ and $B=2\pi \alpha_s \frac{e}{\sqrt{2}} f_{\rho}$. The impact factor $\gamma^{*}_L\rightarrow\rho_L$ has been computed up to twist 2; the next term of the expansion is of twist 4.
It reads \cite{Ginzburg:1985tp}
\begin{equation}
\label{Phi_LL}
\Phi_{\gamma_L\rightarrow\rho_L}(\textbf{k},\textbf{r},Q;\mu^2) = 2 Q B\frac{\delta_{ab}}{2N_c} \int^{1}_{0}dy \:y \, \bar{y} \, \varphi_1(y;\mu^2)P_P(y,\textbf{k},\textbf{r},Q)
\end{equation}
with
\begin{equation}
\label{PLL}
P_P(y,\textbf{k},\textbf{r},Q)=\frac{1}{(y \textbf{r})^2+y\bar{y}Q^2}+\frac{1}{(\bar{y} \textbf{r})^2+y\bar{y}Q^2}-\frac{1}{(\textbf{k}-y \textbf{r})^2+y\bar{y}Q^2}-\frac{1}{(\textbf{k}-\bar{y} \textbf{r})^2+y\bar{y}Q^2}\,.
\end{equation}
 Here $a$ and $b$ are indices of color and $N_c$ is the number of colors.

The first non-vanishing term of the power expansion of the $\gamma^{*}_T\rightarrow\rho_L$ impact
factor has been calculated in Ref.~\cite{Ginzburg:1985tp}. It corresponds to the twist 3 and, in
the limit $t\to t_{\rm{min}}$, reads
\begin{equation}
\label{Phi_LT}
\Phi_{\gamma_T\rightarrow\rho_L}(\textbf{k},\textbf{r},Q;\mu^2) = B\frac{\delta_{ab}}{2N_c} \int^{1}_{0}dy \:(y-\bar{y})\varphi_1(y;\mu^2) \, \eg \cdot \textbf{Q}_P(y,\textbf{k},\textbf{r},Q)
\end{equation}
with
\begin{equation}
\label{QLT}
\textbf{Q}_P(y,\textbf{k},\textbf{r},Q)=\frac{y \textbf{r}}{(y r)^2+y\bar{y}Q^2}-\frac{\bar{y} \textbf{r}}{(\bar{y} r)^2+y\bar{y}Q^2}+\frac{\textbf{k}-y \textbf{r}}{(\textbf{k}-y \textbf{r})^2+y\bar{y}Q^2}-\frac{\textbf{k}-\bar{y}\textbf{r}}{(\textbf{k}-\bar{y}\textbf{r})^2+y\bar{y}Q^2}\,,
\end{equation}
and where
 \beq
\label{def_polT_2d}
 \eg^{\pm}=
\frac{1}{\sqrt{2}}\left(\mp1,- i\right)\,.
\eq
The impact factor for $\gamma^*_T\rightarrow\rho_T$ with the exchanged momentum $\textbf{r} = 0$ is~\cite{Anikin:2009bf}
\beqa
\label{IF_TT}
	&&\Phi_{\gamma_T\rightarrow\rho_T}(\alpha,Q;\mu^2)=\frac{(\epsilon_{\gamma}.\epsilon^{*}_{\rho}) \, 2 B m_{\rho} \delta_{ab}}{2 N_c Q^2}
	\left\{-\int^{1}_{0} dy \frac{\alpha (\alpha +2 y \bar{y})}{y\bar{y}(\alpha+y\bar{y})^2} [(y-\bar{y})\varphi_1^T(y;\mu^2)+\varphi_A^T(y;\mu^2)]\right.\\
&&	+\int^{1}_{0}dy_2\int^{y_2}_{0}dy_1 \frac{y_1\bar{y}_1\alpha}{\alpha+y_1\bar{y}_1}\left[\frac{2-N_c/C_F}{\alpha(y_1+\bar{y}_2)+y_1\bar{y}_2}-\frac{N_c}{C_F}\frac{1}{y_2 \alpha+y_1(y_2-y_1)}\right]M(y_1,y_2;\mu^2)\nonumber \\
&&	\hspace{-.4cm}\left.-\int^{1}_{0}dy_2\int^{y_2}_{0}dy_1 \left[\frac{2+N_c/C_F}{\bar{y}_1}+\frac{y_1}{\alpha+y_1\bar{y}_1}\left(\frac{(2-N_c/C_F)y_1\alpha}{\alpha(y_1+\bar{y}_2)+y_1\bar{y}_2}-2\right)\right. \right.\nonumber \\
&& \hspace{2cm} \left.\left.-\frac{N_c}{C_F}\frac{(y_2-y_1)\bar{y}_2}{\bar{y}_1}\frac{1}{\alpha\bar{y}_1+(y_2-y_1)\bar{y}_2}\right]\!S(y_1,y_2;\mu^2)\right\}\,.\nonumber
\eqa
where $C_F=\frac{N_c^2-1}{2N_c}$ is the quadratic Casimir of the fundamental representation. One can readily check from Eqs.~(\ref{Phi_LL}) and (\ref{Phi_LT})
that $\Phi_{\gamma_L\rightarrow\rho_L}(\textbf{k},\textbf{r},Q;\mu^2)$ and $\Phi_{\gamma_T\rightarrow\rho_L}(\textbf{k},\textbf{r},Q;\mu^2)$
vanish when $\textbf{k} \to 0$ or $\textbf{k} \to \textbf{r}$. Similarly, one can see from
Eq.~( \ref{IF_TT})
that
$\Phi_{\gamma_T\rightarrow\rho_T}(\alpha,Q;\mu^2)$
vanishes in the limit $\alpha \to 0.$



\section{Helicity amplitudes}
\label{sec:helicity_ampl}

\subsection{A phenomenological model for the proton-proton impact factor}
\label{subsec:Model}

To compute ratios of helicity amplitudes, we need a model for the proton-proton impact factor.
A simple  phenomenological model was provided for hadron-hadron scattering in Ref.\cite{Gunion:1976iy}, of the form
\begin{equation}
\label{ProtonIF}
\Phi_{N \to N}(\textbf{k},\textbf{r};M^2)= A \, \delta_{ab}\left[\frac{1}{M^2+(\frac{\textbf{r}}{2})^2}-\frac{1}{M^2+(\textbf{k}-\frac{\textbf{r}}{2})^2}\right]\,.
\end{equation}
 $A$ and $M$  are free parameters that corresponds to the soft scale of the proton-proton impact factor. 
We discuss later the value of $M$. The parameter $A$ has no practical importance for our study since the observables we will be interested in involve  ratios of two scattering amplitudes, which are insensitive to $A$. Note that this impact factor indeed vanishes when $\textbf{k} \to 0$ or $\textbf{r} -\textbf{k} \to 0$ in a minimal way.

This simple model can be interpreted  by assuming that, inside the proton, there exist some typical color-dipole configurations (onia) which will couple to the $\rho-$meson impact factor through a two-gluon exchange.  This impact factor has the same form as a $\gamma^* \to \gamma^*$ impact factor,  with a scale $M^2$ which governs the typical
transverse momentum. Such a model was the basis of the dipole approach of high energy scattering  \cite{Mueller:1993rr} and used successfully for describing deep inelastic scattering at small $x$ \cite{Navelet:1996jx}.

Returning to the discussion given in the Introduction, one can reformulate this model for the proton impact factor at $\textbf{r}=0$ into a simple assumption about the form of $\textbf{k}$ dependence of unintegrated gluon distribution,
\begin{equation}
{\cal F}(x,\textbf{k})\sim \frac{\textbf{k}^2}{\textbf{k}^2+M^2}\, .
\label{form}
\end{equation}

\subsection{Helicity amplitudes $T_{11}$ and $T_{00}$ at $t=t_{min}$ - Comparison with HERA data}
\label{subsec:helicity_ampl_T11_T00}

Let us start from  the impact factor representation at the Born order given by Eq.~(\ref{defImpactRep}).
The helicity amplitude $T_{00}$ is, from Eqs.~(\ref{defImpactRep}, \ref{Phi_LL}, \ref{PLL}, \ref{ProtonIF}),
 \begin{equation}
\label{T00}
 T_{00}= \frac{i s\,  C_F  2 AB}{(2\pi) Q^5}\int^{1}_{0}dy \, \varphi_1(y,\mu^2)\int^{\infty}_{R_1^2}d\alpha  \frac{1}{\alpha^2}\left(\frac{1}{R^2}-\frac{1}{\alpha+R^2}\right)\frac{\alpha}{\alpha+y\bar{y}}\,.
 \end{equation}

In order to obtain  the WW contribution to the $T_{11}$ amplitude, three integrals should be performed. One, over $\alpha$, is related to the transverse momentum of $t-$channel gluons, and two, over $y$ and $u$, are related to DAs; see Eqs.~(\ref{phiAT_WW})
and (\ref{phi1T_WW}).
It is useful to interchange the order of integrals over $\alpha$, $y$, and $u$ in order to  fix a specific model for DAs at the last step when performing the $u$ integration.
This leads to
\beq
\label{T11WW}
T_{11}^{WW}=\frac{i s\,  C_F (\epsilon_{\gamma}.\epsilon^{*}_{\rho}) m_{\rho} 2 A B}{(2\pi) Q^6}\int^{1}_{0}du \frac{\varphi_1(u;\mu^2)}{u}\int^{u}_{0}dy\int^{\infty}_{R_1^2}d\alpha \frac{1}{\alpha^2}\left(\frac{1}{R^2}-\frac{1}{\alpha+R^2}\right) \frac{\alpha(\alpha+2y\bar{y})}{(\alpha+y\bar{y})^2}
\eq
where  $R^2=\frac{M^2}{Q^2}$. We also introduced a cutoff $R_1^2 = \frac{\lambda^2}{Q^2}$ on the integral over $\alpha$\,, where $\lambda$ is the cutoff on $\left|\textbf{k}\right|$. This cutoff allows us to see how soft gluons contribute to the amplitude.

The genuine contribution is 
\beqa
\label{T11gen}
&&T_{11}^{gen} = \frac{i s C_F (\epsilon_{\gamma}.\epsilon^{*}_{\rho}) m_{\rho} AB}{ (2\pi) Q^6}\int^{\infty}_{R_1^2}d\alpha \left\{\frac{1}{R^2}-\frac{1}{\alpha+R^2}\right\} \left\{-\int^{1}_{0} dy \frac{\alpha (\alpha +2 y \bar{y})}{y\bar{y}(\alpha+y\bar{y})^2} [(y-\bar{y})\varphi^{gen}_{1T}(y;\mu^2)+\varphi^{gen}_{AT}(y;\mu^2)] \right. \nonumber\\
	&&+\int^{1}_{0}dy_2\int^{y_2}_{0}dy_1 \frac{y_1\bar{y}_1\alpha}{\alpha+y_1\bar{y}_1}\left[\frac{2-N_c/C_F}{\alpha(y_1+\bar{y}_2)+y_1\bar{y}_2}-\frac{N_c}{C_F}\frac{1}{y_2 \alpha+y_1(y_2-y_1)}\right]M(y_1,y_2;\mu^2)\nonumber\\
	&&\hspace{-.3cm}\left.-\int^{1}_{0}dy_2\int^{y_2}_{0}dy_1 \left[\frac{2+N_c/C_F}{\bar{y}_1}+\frac{y_1}{\alpha+y_1\bar{y}_1}\left(\frac{(2-N_c/C_F)y_1\alpha}{\alpha(y_1+\bar{y}_2)+y_1\bar{y}_2}-2\right)\right.\right. \nonumber \\
&&\hspace{2cm} \left. \left.-\frac{N_c}{C_F}\frac{(y_2-y_1)\bar{y}_2}{\bar{y}_1}\frac{1}{\alpha\bar{y}_1+(y_2-y_1)\bar{y}_2}\right]S(y_1,y_2;\mu^2)\right\}\,.
\eqa
For convenience, we define $I_1(y;R^2,R_1^2)$, $I_2(y_1,y_2;R^2,R_1^2),$ and $I_3(y_1,y_2;R^2,R_1^2)$ as the integrands after integration over $\alpha$:
\beqa
\label{def_I1}
I_1(y;R^2,R_1^2) &=& \int^{\infty}_{R_1^2}d\alpha \left(\frac{1}{R^2}-\frac{1}{\alpha+R^2}\right)\frac{\alpha (\alpha +2 y \bar{y})}{y\bar{y}(\alpha+y\bar{y})^2}\,,
\\
\label{def_I2}
I_2(y_1,y_2;R^2,R_1^2)&=&
\int^{\infty}_{R_1^2}d\alpha \left\{\frac{1}{R^2}-\frac{1}{\alpha+R^2}\right\}
\frac{y_1\bar{y}_1\alpha}{\alpha+y_1\bar{y}_1}\left[\frac{2-N_c/C_F}{\alpha(y_1+\bar{y}_2)+y_1\bar{y}_2}-\frac{N_c}{C_F}\frac{1}{y_2 \alpha+y_1(y_2-y_1)}\right]\,,
\\
\label{def_I3}
I_3(y_1,y_2;R^2,R_1^2)&=&
\int^{\infty}_{R_1^2}d\alpha \left\{\frac{1}{R^2}-\frac{1}{\alpha+R^2}\right\}
\left[\frac{2+N_c/C_F}{\bar{y}_1}+\frac{y_1}{\alpha+y_1\bar{y}_1}\left(\frac{(2-N_c/C_F)y_1\alpha}{\alpha(y_1+\bar{y}_2)+y_1\bar{y}_2}-2\right)\right. \nonumber \\
&&\hspace{2cm}  \left.-\frac{N_c}{C_F}\frac{(y_2-y_1)\bar{y}_2}{\bar{y_1}}\frac{1}{\alpha\bar{y}_1+(y_2-y_1)\bar{y}_2}\right]\,,
\eqa
such that (\ref{T11gen}) can be expressed as (removing the variables $R^2$ and $R_1^2$ for simplicity)
\beqa
\label{T11_gen1}
T_{11}^{gen}&=&  \frac{i s \, C_F (\epsilon_{\gamma}.\epsilon^{*}_{\rho}) m_{\rho} AB}{ (2\pi) Q^6}\left\{-\int^{1}_{0} dy \, I_1(y) [(y-\bar{y})\varphi^{gen}_{1T}(y;\mu^2)+\varphi^{gen}_{AT}(y;\mu^2)]\right.\nonumber \\
	&&\left.+\int^{1}_{0}dy_2\int^{y_2}_{0}dy_1 I_2(y_1,y_2) M(y_1,y_2;\mu^2)
	-\int^{1}_{0}dy_2\int^{y_2}_{0}dy_1 I_3(y_1,y_2) S(y_1,y_2;\mu^2)\right\}\,,
\eqa
which, using the symmetry property $S(y_1,y_2;\mu^2)= -M(\bar{y}_2,\bar{y}_1;\mu^2)$, turns into
\begin{multline}
\label{T11_gen2}
T_{11}^{gen}= \frac{i s C_F(\epsilon_{\gamma}.\epsilon^{*}_{\rho}) m_{\rho} AB}{(2\pi) Q^6}\left\{-\int^{1}_{0} dy I_1(y) [(y-\bar{y})\varphi^{gen}_{1T}(y;\mu^2)+\varphi^{gen}_{AT}(y;\mu^2)] \right.\\
	\hspace{-4cm}\left.+\int^{1}_{0}dy_2\int^{y_2}_{0}dy_1 (I_2(y_1,y_2)+I_3(\bar{y}_2,\bar{y}_1) )M(y_1,y_2;\mu^2)\right\}
\end{multline}
with
\begin{multline}
\label{sumI2I3}
I_2(y_1,y_2)+I_3(\bar{y}_2,\bar{y}_1) = \left(2-\frac{N_c}{C_F}\right)\int^{\infty}_{R_1^2}d\alpha \frac{1}{R^2(\alpha+R^2)(\alpha(y_1+\bar{y}_2)+y_1\bar{y}_2)}\left(\frac{\bar{y}_2^2}{\alpha+y_2\bar{y}_2}+\frac{y_1\bar{y}_1}{\alpha+y_1\bar{y}_1}\right)\\
\hspace{-4cm}+\frac{N_c}{C_F} \int^{\infty}_{R_1^2}d\alpha \frac{1}{R^2(\alpha+R^2)(\alpha+y_1\bar{y}_1)(\alpha y_2+y_1(y_2-y_1))} +\frac{2}{y_2} \int^{\infty}_{R_1^2}d\alpha \frac{1}{R^2(\alpha+R^2)(\alpha+y_2\bar{y}_2)} \,.
\end{multline}

Combining the results (\ref{T00}) with (\ref{T11WW}) and (\ref{T11_gen2}), the ratios $T_{11}^{WW}/T_{00}$ and $T_{11}^{gen}/T_{00}$ read
\begin{equation}
\label{res_T11_sur_T00_WW}
\frac{T_{11}^{WW}}{T_{00}}(Q,M,\lambda) = \displaystyle - \frac{\displaystyle m_{\rho}}{\displaystyle Q} \frac{\displaystyle \int^{1}_{0}dv\, \varphi_1(v;\mu^2)\int^{1}_{0}dx\int^{\infty}_{R_1^2}d\alpha \frac{\alpha+2xv(1-xv)}{\alpha(\alpha+xv (1-xv))^2} \left(\frac{1}{R^2}-\frac{1}{\alpha+R^2}\right)}{\displaystyle\int^{1}_{0}dy \, \varphi_1(y,\mu^2)\int^{\infty}_{R_1^2}  \frac{d\alpha}{\alpha(\alpha+y\bar{y})} \left(\frac{1}{R^2}-\frac{1}{\alpha+R^2}\right)}
\end{equation}
where we took into account that $\epsilon_{\gamma}.\epsilon^{*}_{\rho}=-\eg_{\gamma} \cdot \boldsymbol{\epsilon}^{*}_{\rho} =-1$, and
\begin{equation}
\label{res_T11_sur_T00_gen}
\frac{T_{11}^{gen}}{T_{00}} \!=\!  \frac{\displaystyle m_\rho}{\displaystyle 2Q}\frac{\displaystyle \int^{1}_{0} dy \, I_1(y) [(y-\bar{y})\varphi^{gen}_{1T}(y;\mu^2)+\varphi^{gen}_{AT}(y;\mu^2)]
-\!\int^{1}_{0}dy_2\int^{y_2}_{0}dy_1 (I_2(y_1,y_2)+I_3(\bar{y}_2,\bar{y}_1) )M(y_1,y_2;\mu^2)}{\displaystyle\int^{1}_{0}dy \, \varphi_1(y;\mu^2)\int^{\infty}_{R_1^2}d\alpha  \frac{1}{\alpha^2}\left(\frac{1}{R^2}-\frac{1}{\alpha+R^2}\right)\frac{\alpha}{\alpha+y\bar{y}}}\,.
\end{equation}
The integration is performed analytically over $\alpha$ and numerically over the variables left as, for example, $y$ for $T_{00}$, $x$, $v$ for $T_{11}^{WW}$, and $y_1$, $y_2$ for $T_{11}^{gen}$\,.
The measured ratio $T_{11}/T_{00}$ is conventionally  defined  \cite{:2010dh} to have the {\em opposite}  sign with respect to
Eqs.~(\ref{res_T11_sur_T00_WW}) and (\ref{res_T11_sur_T00_gen}), in order to ensure the usual matrix summation in the definition of the density matrix.

\begin{figure}[h!]
 \includegraphics[width=0.5\textwidth]{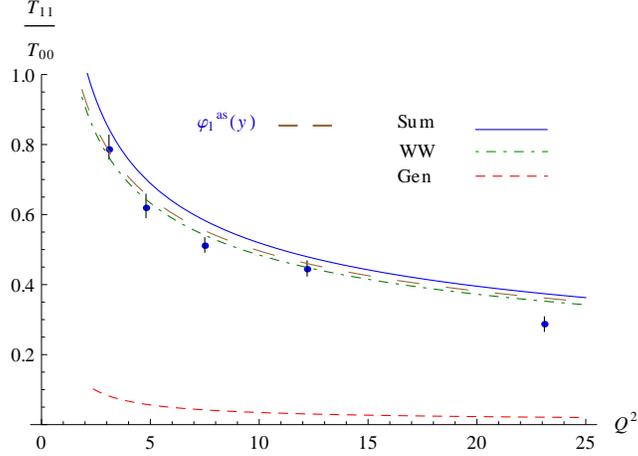}
\caption{The WW contribution $T_{11}^{WW}/T_{00}$ in green (dash-dotted line), the genuine contribution $T_{11}^{gen}/T_{00}$ in red (dashed line), and the sum of the two contributions in blue (solid line), at $M=1$ GeV and $\lambda=0$ GeV, as functions of the virtuality of the photon.
The brown (long-dashed) curve is the contribution based on the asymptotic DA of the $\rho$ meson,
$\varphi_1(y,\mu^2=\infty)=\varphi_{1}^{as}(y)=6 y (1-y).$
 Our results are compared
  with the experimental data from H1 \cite{Aaron:2009xp}. The experimental errors are taken to be the quadratic sum of statistical and systematical errors.}
	\label{T_11_00_WW_gen_rho}
\end{figure}

In  Fig.~\ref{T_11_00_WW_gen_rho} we show different contributions  to the ratio $T_{11}/T_{00}$ as a function of $Q^2$ and for
typical values of nonperturbative parameters $M$ and $\lambda$. Unless specified, we take as a factorization scale $\mu=Q$. Note that the factorization scale only appears in the ratio of the amplitudes through the DAs and the coupling constants. We see that the WW contribution dominates over the genuine one.
For illustration, we also show this ratio using the asymptotic $\varphi_1^{as}=6y\bar y$ DA, which corresponds to $\mu \to \infty$, which is identical to the WW contribution since the genuine twist-3  contribution vanishes in this limit. The small difference between
this last curve and the solid one shows a weak dependence of this ratio on the factorization scale $\mu$.

\begin{figure}[h!]
	\hspace{-.5cm}
\includegraphics[width=0.44\textwidth]{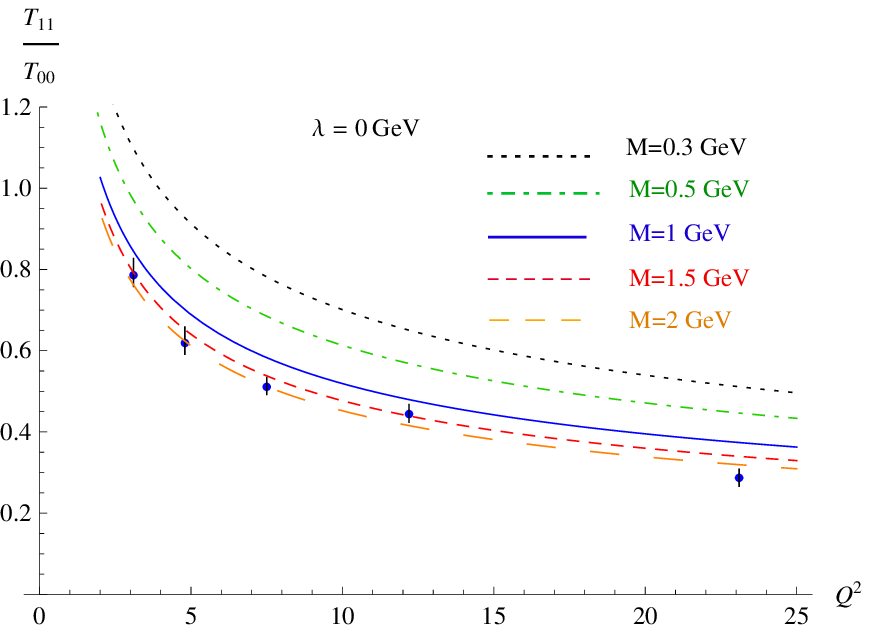} \hspace{1cm} \includegraphics[width=0.44\textwidth]{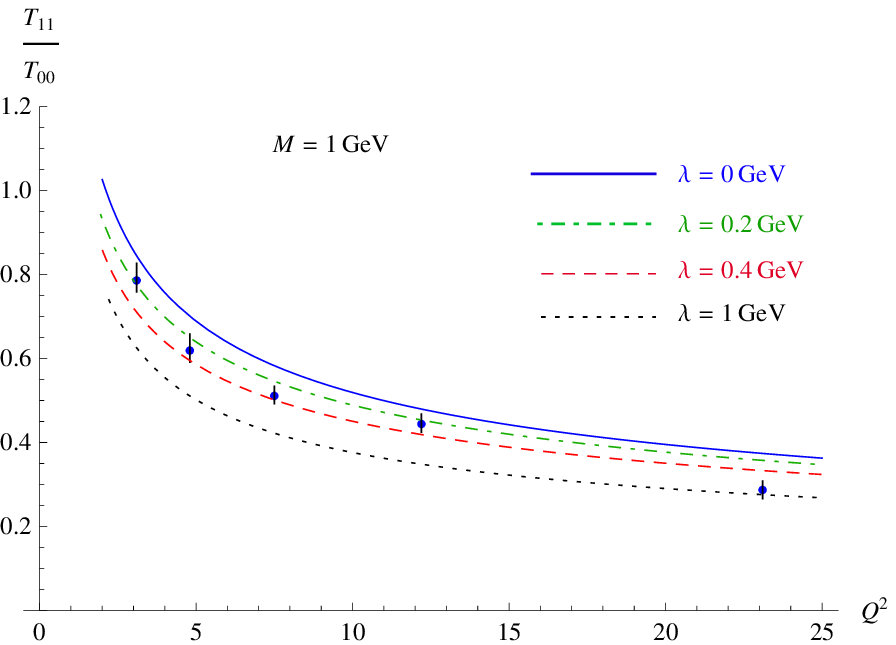}
\caption{Predictions for the ratio $T_{11}/T_{00}$ as a function of $Q^2$, compared  to the experimental data from H1 \cite{Aaron:2009xp}. The experimental errors are taken to be the quadratic sum of statistical and systematical errors. Left panel: Fixed $\lambda=0$ GeV cutoff and various values for $M$. Right panel: Fixed scale $M=1$ GeV, and various values of the cutoff $\lambda$.}
	\label{T_11_00_param}
\end{figure}

The two parameters $\lambda$ and $M$ have different physical meanings. $M$ is the typical nonperturbative hadronic scale, while $\lambda$ is the minimal virtuality of gluons, which should be bigger than $\Lambda_{QCD}$ for consistency of  our perturbative approach.
From Fig.~\ref{T_11_00_param}~(left panel), we see that our predictions are stable for $M$ in the range 1-2 GeV. The data, when compared with our model, with $\mu=Q$, favor a value of  $M$ of the order of 1-2 GeV but exclude a very small value around $\Lambda_{QCD}.$
From
Fig.~\ref{T_11_00_param}~(right panel), we see that for $\lambda$ around $\Lambda_{QCD}$, our results are very close to the experimental data and rather stable, whereas for $\lambda=1$ GeV, i.e. significantly larger than $\Lambda_{QCD} \simeq 220$ MeV in the $\overline{\rm MS}$ scheme, they notably deviate from the data.
Let us stress that the fact that our estimate provides the correct sign  for the ratio $T_{01}/T_{00}$ when compared to H1 data is a nontrivial success of our approach.

\begin{figure}[h!]
 \includegraphics[width=0.5\textwidth]{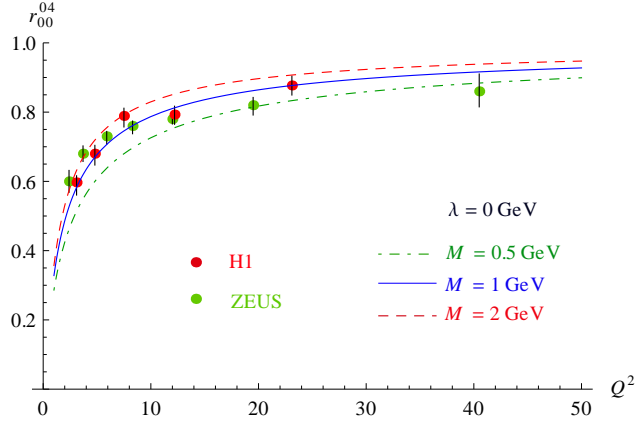}
\caption{The spin density matrix element $r^{04}_{00}$ as a function of $Q^2$ for $M=0.5$ GeV in green (dash-dotted line), $M=1$ GeV in blue (solid line), and $M=2$ GeV in red (dashed line), and for $\lambda=0$ GeV. Our results are compared
  with the experimental data from  ZEUS \cite{Chekanov:2007zr} and H1 \cite{Aaron:2009xp}. The experimental errors are taken to be the quadratic sum of statistical and systematical errors.}
	\label{r04}
\end{figure}

In  Fig.~\ref{r04} we show the result of our calculations for the spin density matrix element $r^{04}_{00}$ as a function of $Q^2$ and for
typical values of the nonperturbative parameter $M$ and for $\lambda=0$ GeV.
This observable allows a comparison of our prediction with the whole set of HERA data\footnote{We predict ratios of amplitudes, while ZEUS made available the spin density matrix elements;
H1 extracted both spin density matrix elements and ratios of amplitudes.}. Note that our amplitudes are evaluated at $t=t_{min}$ while experimental data are integrated over some $t$ range but dominated by very small values of $t$. At $t=t_{min}$,   $r^{04}_{00}$ only depends on the ratio $x_{11}=|T_{11}|/|T_{00}|$ through
\beq
\label{r0_04_ratioT}
r^{04}_{00}=\frac{\varepsilon}{\varepsilon+x_{11}^2}\,,
\eq
where $\varepsilon$ is the photon polarization parameter $\varepsilon \simeq (1-y)/(1-y+y^2/2)$. For H1 $\langle \varepsilon \rangle
=0.98$ and for ZEUS $\langle \varepsilon \rangle
=0.996.$

For $t \neq t_{min}$, $r^{04}_{00}$ only slightly depends on the $s$-channel helicity violating amplitudes $T_{01}$,  $T_{10}$, and $T_{1-1}$. Experimental data are dominated by $|t - t_{min}| \leq 0.4$ GeV$^2$, for which the only significant amplitudes are \\
 $|T_{00}| >|T_{11}|  > |T_{01}|$. The exact relation reads
\beq
\label{r0_04_ratioT-detail}
r^{04}_{00}=\frac{\varepsilon+x_{01}^2}{x_{11}^2+\varepsilon+x_{01}^2 + x_{1-1}^2+2 \varepsilon \, x_{10}^2}\,
\eq
where $x_{ij}=|T_{ij}|/|T_{00}|$. The expression (\ref{r0_04_ratioT-detail})
simplifies when one neglects the contributions of $x_{1-1}$ and $x_{10}$, and expands the remaining expression up to the first power of $\frac{x_{01}^2}{\varepsilon +x_{11}^2}$. In this way, we get
\beq
\label{simp_r0_04_ratioT-detail}
r^{04}_{00}\approx \frac{1}{\varepsilon+x_{11}^2}\left(\varepsilon+ \frac{x_{01}^2 x_{11}^2}{\varepsilon+x_{11}^2} \right)
\,.
\eq
Based on H1 and ZEUS data, the effect of the second term in the parenthese of Eq.~(\ref{simp_r0_04_ratioT-detail}) is below 1 $\%$, so up to this accuracy the expression (\ref{simp_r0_04_ratioT-detail}) is equivalent to the formula (\ref{r0_04_ratioT}).

\subsection{Helicity amplitudes $T_{00}$ and $T_{01}$ for $t \neq t_{min}$}
\label{subsec:helicity_ampl_T00_T01}

\begin{figure}[htbp]

\hspace{-.5cm}
		\includegraphics[width=0.44\textwidth]{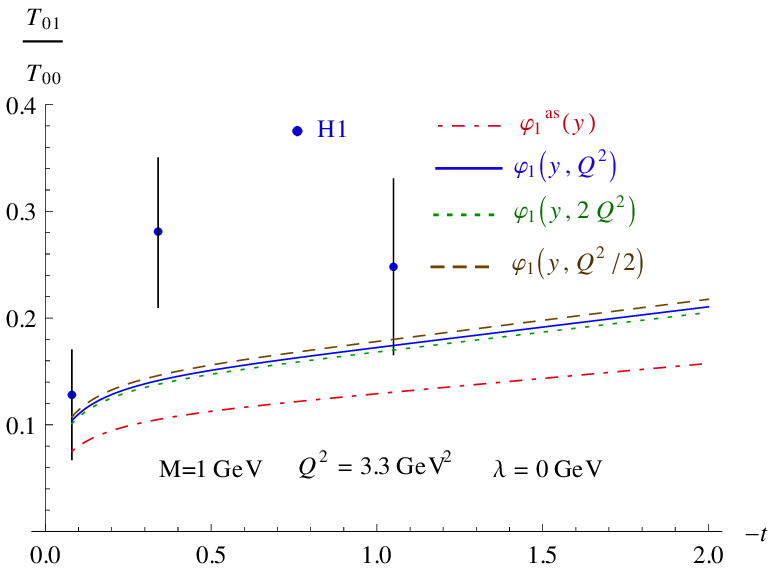}
\hspace{.5cm}
\includegraphics[width=0.44\textwidth]{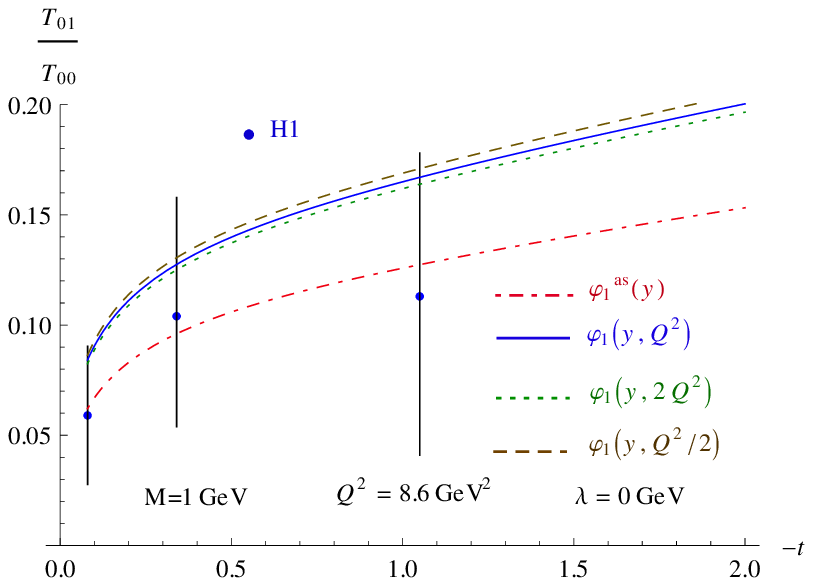}
	\caption{Predictions for the ratios $T_{01}/T_{00}$ as a function of $\left|t\right|$  for $M=1$ GeV and $\lambda=0$ GeV, for different values of the factorization scale $\mu^2$, compared with H1 data \cite{Aaron:2009xp}:
the blue  (solid) line is for $\varphi_{1}(y,\mu^2=Q^2)$, the green (dotted) line is for $\varphi_{1}(y,\mu^2=2 Q^2)$, the brown (dashed) line is for $\varphi_{1}(y,\mu^2=Q^2/2)$, and the red (dashed) line is for $\varphi_1(y,\mu^2=\infty)=\varphi_{1}^{as}(y)=6 y (1-y)$. The experimental errors are taken to be the quadratic sum of statistical and systematical errors. Left panel:
$Q^2=3.3$ GeV$^2$. Right panel: $Q^2=8.6$ GeV$^2$.}
	\label{mu_dep_T01}
\end{figure}

The H1 data show that the spin-flip amplitude $T_{01}$ is nonzero, showing an explicit $s-$channel
helicity violation.  Besides,
this amplitude vanishes  when the squared momentum exchanged by the proton $t = -\textbf{r}^2$ is zero.
We start with the generalization of Eq.~(\ref{T00}) for  $t \neq t_{min}$:
\beqa
\label{defT00_general}
T_{00}&=&\frac{is \, C_F 2\, Q AB}{(2\pi)^2 (M^2+(\rg/2)^2)} \int^{1}_{0}dy\:y\bar{y} \, \varphi_1(y; \mu^2)\int \frac{d^2 \textbf{k}}{\kg^2 (\kg-\rg)^2}
\frac{(\kg-\rg/2)^2-(\rg/2)^2}{(\kg-\rg/2)^2+M^2} \nonumber
\\
&&
\times \left\{\frac{1}{(y \rg)^2+y\bar{y} \,Q^2}+\frac{1}{(\bar{y}\rg)^2+y\bar{y} \,Q^2}-\frac{1}{(\kg-y \, \rg)^2+y\bar{y} \, Q^2}-\frac{1}{(\kg-\bar{y} \rg)^2+y\bar{y} \,Q^2}\right\} \,.
\eqa
Similarly,
\beqa
\label{defT01_general}
T_{01}&=&\frac{is \, C_F 2\, Q AB}{(2\pi)^2 (M^2+(\rg/2)^2)} \int^{1}_{0}dy\:(y-\bar{y}) \, \varphi_1(y; \mu^2)\int \frac{d^2 \textbf{k}}{\kg^2 (\kg-\rg)^2}
\frac{(\kg-\rg/2)^2-(\rg/2)^2}{(\kg-\rg/2)^2+M^2} \nonumber
\\
&&
\times \left\{\frac{y \, \rg \cdot \eg}{(y \rg)^2+y\bar{y} \,Q^2}-\frac{\bar{y} \, \rg \cdot \eg}{(\bar{y}\rg)^2+y\bar{y} \,Q^2}+\frac{(\kg-y \, \rg) \cdot \eg}{(\kg-y \, \rg)^2+y\bar{y} \, Q^2}-\frac{(\kg-\bar{y} \, \rg) \cdot \eg}{(\kg-\bar{y} \rg)^2+y\bar{y} \,Q^2}\right\} \,.
\eqa
The $k_T$ integrations in Eqs.~(\ref{defT00_general}, \ref{defT01_general}) are performed in two different ways, presented in detail in Appendix \ref{subsec:T00_T01_general}:

\begin{itemize}
\item the integrations over $k_T$ are performed, without any infrared cutoff, partially analytically through a residue method (see Appendix \ref{subsubsec:No_IR_cutoff}),

\item the integrations over $k_T$ are performed, with an infrared cutoff, fully numerically through
triangulation coordinates centered at the pole of the two $t-$channel gluons (see Appendix \ref{subsubsec:IR_cutoff}).
\end{itemize}
Figure \ref{mu_dep_T01} shows the dependence of the ratio $T_{01}/T_{00}$ on the choice of the factorization scale $\mu$ for $M=1$ GeV and $\lambda=0$ GeV. For completeness, we also show the predictions based on the asymptotic DAs. We see that for factorization scales around $\mu^2=Q^2$ our results are rather insensitive to its values. Nevertheless, the ratio $T_{01}/T_{00}$ seems to be more sensitive to this scale than the ratio $T_{11}/T_{00}$.

\begin{figure}[htbp]
	\hspace{-.5cm}
		\includegraphics[width=0.44\textwidth]{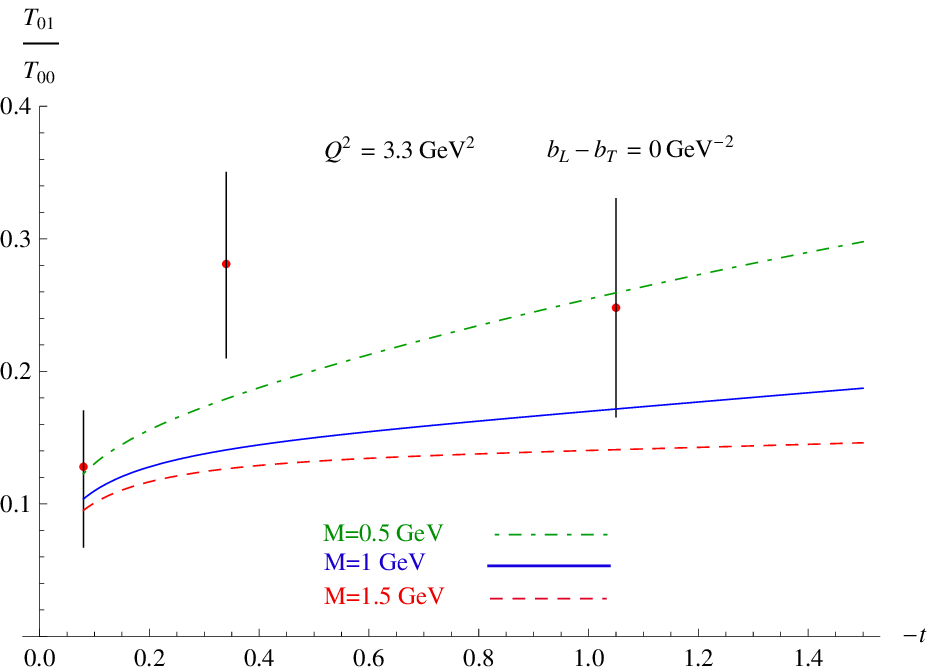}
\hspace{.5cm}
\includegraphics[width=0.44\textwidth]{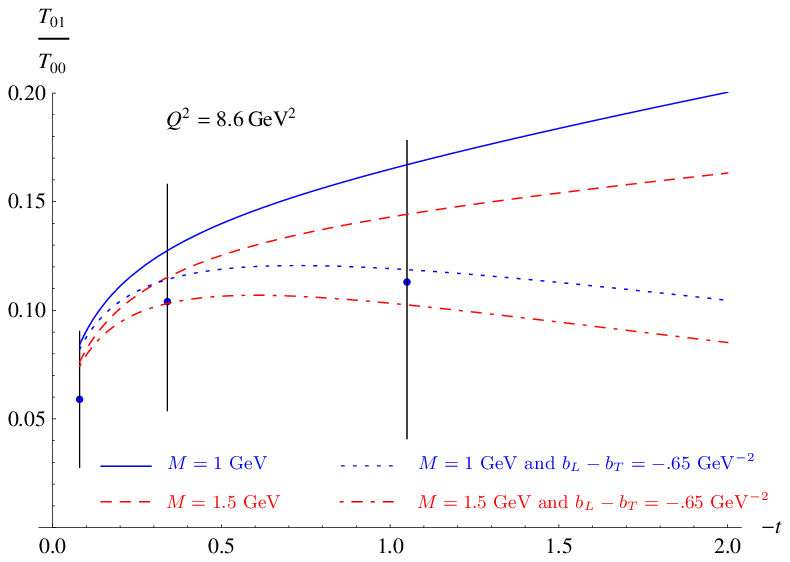}
	\caption{Predictions for the ratio $T_{01}/T_{00}$ as a function of $\left|t\right|$  for $\lambda=0$ GeV, for various values of $M$, compared with H1 data \cite{Aaron:2009xp}. The experimental errors are taken to be the quadratic sum of statistical and systematical errors. Left panel:  $Q^2=3.3$ GeV$^2$. Right panel: $Q^2=8.6$ GeV$^2$. }
	\label{fig:T01delta_b}
\end{figure}

Our predictions are based on perturbative QCD and therefore, at small $t$, can only lead to a powerlike or logarithmic $t$ dependence. On the other hand, it is well known that data exhibit an exponentially falling $t$ distribution, as has been seen both by the H1 \cite{Aaron:2009xp} and ZEUS \cite{Chekanov:2007zr} collaborations. Therefore, we have studied the effect of multiplying our predictions for the amplitudes by a factor $e^{-b_i \, |t-t_{min}|/2}$, where $b_i$ $(i=L,T)$ corresponds to
 $\rho$ electroproduction from $\gamma^*_L$ or $\gamma^*_T\,.$
 H1 measured values of $b_L$ and $b_L-b_T$ \cite{Aaron:2009xp}. The measured values for the latter are $b_L -b_T = -0.03 \pm 0.27^{+0.19}_{-0.17}$ GeV$^{-2}$ (for  $\langle Q^2\rangle=3.3 \ {\rm GeV}^2$) and
$b_L -b_T = -0.65 \pm 0.14^{+0.41}_{-0.51}$ GeV$^{-2}$ (for  $\langle Q^2\rangle=8.6 \ {\rm GeV}^2$).
Here we present our results in Fig.~\ref{fig:T01delta_b}.
 One can see in the right panel of Fig.~\ref{fig:T01delta_b}
that the precision of the data 
for the $T_{01}/T_{00}$ ratio does not permit
 us to discriminate between a zero value for the difference of the transverse and the
longitudinal slope parameters, $b_L-b_T$, and a nonzero value of this difference, as
measured by H1 at higher values of $Q^2$.

Thus our estimate provides the correct sign and order of magnitude for the ratio $T_{01}/T_{00}$ when compared to H1 data for $M$ of the order of 1 GeV
in the whole range of $ \langle -t \rangle < 1.08 $
GeV$^2$.

\def\sca{.6}


\begin{figure}[htbp]
	
\hspace{-.5cm}
		\includegraphics[width=0.44\textwidth]{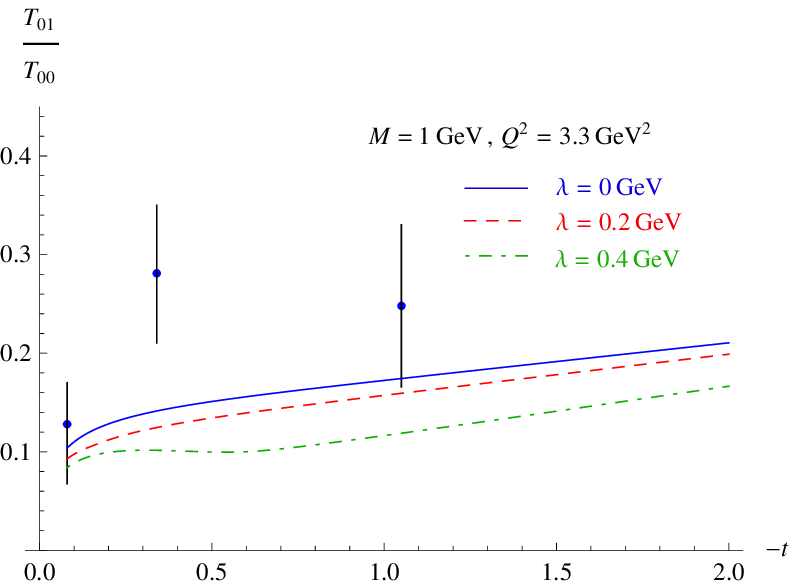}
\hspace{.5cm}
\includegraphics[width=0.44\textwidth]{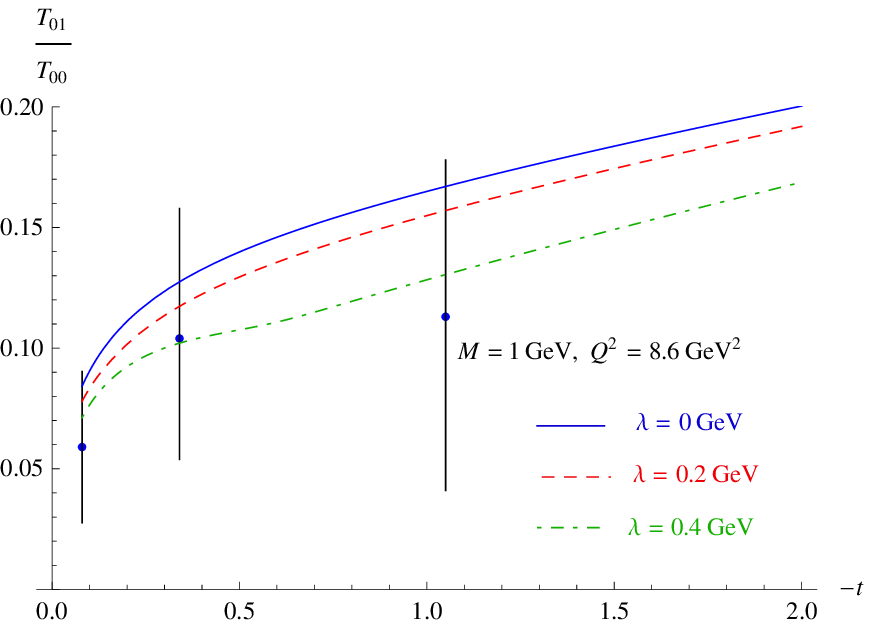}
	\caption{Predictions for the ratios $T_{01}/T_{00}$ as a function of $\left|t\right|$  for $M=1$ GeV, for different values of $\lambda$, compared with H1 data \cite{Aaron:2009xp}:
the blue  (solid) line is for $\lambda=0$ GeV,  the red (dashed) line is for $\lambda=0.2$ GeV, and the green (dash-dotted) line is for $\lambda=0.4$ GeV. The experimental errors are taken to be the quadratic sum of statistical and systematical errors. Left panel:
$Q^2=3.3$ GeV$^2$. Right panel: $Q^2=8.6$ GeV$^2$.}
	\label{fig:3pt3_8pt6_M1_k_10_la0red_la0pt3green}
\end{figure}

%

For completeness, as we did for the ratio $T_{11}/T_{00}$, we also display in Fig.~\ref{fig:3pt3_8pt6_M1_k_10_la0red_la0pt3green} the effect of varying the cutoff $\lambda$ on $k_T$ for the ratio
$T_{01}/T_{00}$. Again, the prediction does not change significantly when $\lambda$ is around $\Lambda_{QCD}$.
One obtains the same kind of values for $M$ and $\lambda$ when comparing with the data for the two ratios
$T_{11}/T_{00}$ and $T_{01}/T_{00}$.
 However, due to a lack of precision of the data for the ratio $T_{01}/T_{00}$, the parameters $M$ and $\lambda$ are mainly
constrained by the ratio $T_{11}/T_{00}$.

\section{Conclusion}
\label{sec:conclusion}

We have evaluated the ratios $T_{11}/T_{00}$ and
$T_{01}/T_{00}$ within $k_T$ factorization, which we compared with recent H1 data. We got fairly good
agreement with reasonable values of the two nonperturbative parameters $\lambda$ and $M$ involved in our description. In particular, we found rather weak sensitivity of the obtained helicity amplitudes to the region of small $t$-channel gluon momenta (small values of $\lambda$). This justifies our approach to the $\Phi^{\gamma^{*}(\lambda_{\gamma}) \rightarrow \rho(\lambda_{\rho})}(\textbf{k}^2,Q^2)$ impact factor,  based on the assumption about the dominance of the scattering of small transverse-size quark-antiquark and quark-antiquark-gluon colorless states on the target, which includes its calculation using QCD twist expansion in terms of meson DAs.

Our calculation of the transverse amplitude involves both WW and genuine twist-3 contributions. It turns out that
with the input for coupling constants in DAs of $\rho$ mesons, determined from the QCD sum rules \cite{Ball:1998sk}, the WW contribution strongly dominates these two observables. Besides, we found rather weak dependence of the WW contribution on the shape of meson twist-2 DAs. In other words, we found that the $T_{11}/T_{00}$ amplitude ratio turned out to be not very sensitive to the physics encoded in the meson DAs. This opens a possibility to constrain the other important quantity which enters the $k_T$-factorization formalism, the unintegrated gluon distribution ${\cal F}(x,\textbf{k})$.
In the present study we use the very simple ansatz (\ref{form}) for the $\textbf{k}$ shape of ${\cal F}(x,\textbf{k})$.
Having rather precise data for the $T_{11}/T_{00}$ amplitude ratio, it would be very interesting to test different approaches to unintegrated gluon distribution present in the literature, calculating this ratio in $k_T$ factorization.

The lack of  precision for the present data for the ratio $T_{01}/T_{00}$
does not allow us to fix precisely the nonperturbative parameters in our description. Nevertheless, their values  ($M$ being of the order of 1-2 GeV and $\lambda$ being of the order of 0-0.2 GeV) extracted from the $T_{11}/T_{00}$ ratio analysis and used for the prediction of the $T_{01}/T_{00}$ ratio do not contradict data.

Other scattering amplitude ratios have also been measured and  should be confronted by a $k_T$-factorization approach. This
requires nontrivial analytical calculations  for $t \neq t_{min}$ of the twist-3 amplitudes (which was not needed for the ratio
$T_{01}/T_{00}$) which is a hard task, since it involves, in particular, the computation of the $\gamma^*_L \to \rho_T$ impact factor. This deserves a separate study.

Data also exist for $\phi$ leptoproduction. In this case quark-mass effects should be taken into account, in particular, because this allows the transversely polarized $\phi$ to couple through its chiral-odd twist-2 DA.
The fact that the ratio $T_{11}/T_{00}$ is not the same (after trivial mass rescaling) for $\rho$ and $\phi$ mesons points to the importance of this effect.
This is also beyond the scope of our present study, but may open an interesting  way for accessing chiral-odd DAs.
%

The BFKL resummation effects have not been included. Although they were    expected \cite{Ginzburg:1985tp, Ginzburg:1988zy} to be rather dramatic  at the level of scattering amplitudes, as was confirmed by a leading logarithmic $x$ analysis of HERA data \cite{Forshaw:2001pf, Enberg:2003jw, Poludniowski:2003yk},
we expect that for the ratios of amplitudes considered here, they should be rather moderate. Also, the
next-to-leading order effects - both on the evolution and on the impact factor - should be studied. 
%
%
%
%

On the experimental side, 
the future Electron-Ion Collider with a high center-of-mass energy and high luminosities, as well as the International Linear Collider, will hopefully open  the opportunity  to study in more detail the hard diffractive production of mesons \cite{Pire:2005ic,Enberg:2005eq, Ivanov:2005gn,Pire:2006ik,Ivanov:2006gt,Segond:2007fj,Caporale:2007vs,Segond:2008ik}.

\vskip.2in \noindent
\acknowledgments

\noindent
We acknowledge a very useful discussion with Sergey Manaenkov. This work is supported in part by RFBR Grants No. 09-02-00263, No. 11-02-00242,
No. NSh-3810.2010.2,
and by the Polish Grant No. N202 249235 and the French-Polish Collaboration Agreement Polonium.


\appendix

\section{Distribution amplitudes}
\label{subsec:DA}

Distribution amplitudes are normalized as 
\beq
\label{norm_DA}
\int^{1}_{0} dy \, \varphi_1(y;\mu^2)=1 \,, \qquad
\int^{1}_{0} dy \,  \varphi_3(y;\mu^2)=1 \,, \qquad
 \int^{1}_{0} dy \, (y-\bar{y})\varphi_A(y;\mu^2)=\frac{1}{2}\,.
 \eq
 The following symmetry properties are derived from $\C$-parity analysis:
 \beqa
\label{sym_A}
&& \varphi_1(\bar{y};\mu^2)=\varphi_1(y;\mu^2) \,, \qquad
  \varphi_3(\bar{y};\mu^2)=\varphi_3(y;\mu^2) \,, \qquad
   \varphi_A(\bar{y};\mu^2)=-\varphi_A(y;\mu^2)\,, \nonumber \\
&& \varphi_1^T(\bar{y};\mu^2)=-\varphi_1^T(y;\mu^2) \,, \qquad
  \varphi_A^T(\bar{y};\mu^2)=\varphi_A^T(y;\mu^2) \,, \nonumber \\
&&  B(y_1,y_2;\mu^2)=-B(\bar{y}_2,\bar{y}_1;\mu^2)\,, \qquad
  D(y_1,y_2;\mu^2)=D(\bar{y}_2,\bar{y}_1;\mu^2)\,.
  \eqa
Then, the functions $M$ and $S$ defined in Eq.~(\ref{defSM}) satisfy
\begin{equation}M(y_1,y_2;\mu^2)=-S(\bar{y}_2,\bar{y}_1;\mu^2)\,.
\end{equation}

\section{Evolutions of DAs as a function of $Q$}
\label{subsec:DA_evolution}

The parameters entering the DAs at $\mu=1$ GeV and their evolution equations  are given in Ref.~\cite{Ball:1998sk}. In Table~\ref{table:DAs} we recall their values for   the $\rho$ meson at $\mu_0=1$ GeV.
\begin{table}[htbp]
	\centering
 \begin{tabular}{|l|c|r|}
\hline
 $\alpha_s$ & 0.52\\ \hline
& \\
$\omega^{A}_{\{1,0\}}$ & -2.1\\ \hline
&\\
$\omega^{V}_{[0,1]}$ & 28/3\\ \hline
&\\
$a_{2,\rho}$ & 0.18 $\pm$ 0.10\\ \hline
&\\
$m_\rho \,f^{A}_{3\rho}$ & $0.5-0.6\; 10^{-2} \,{\rm GeV}^2$\\ \hline
&\\
$m_\rho \, f^{V}_{3\rho}$ & $0.2 \:10^{-2} \,{\rm GeV}^2$\\ \hline
&\\
$\zeta^{A}_{3\rho}$ & 0.032\\ \hline
&\\
$\zeta^{V}_{3\rho}$ & 0.013\\ \hline
\end{tabular}
\caption{Coupling constants and Gegenbauer coefficients entering the $\rho-$meson DAs, at the scale $\mu_0=1$ GeV. Note that in Ref.~\cite{Ball:1998sk} the normalization are such that $f^{V,A}_{3\rho\, \mbox{\cite{Ball:1998sk}}}=m_\rho \, f^{V,A}_{3\rho\, [{\rm here}]}$.}
\label{table:DAs}
\end{table}

\begin{figure}[htbp]
	\begin{tabular}{ccc}
		\includegraphics[width=0.325\textwidth]{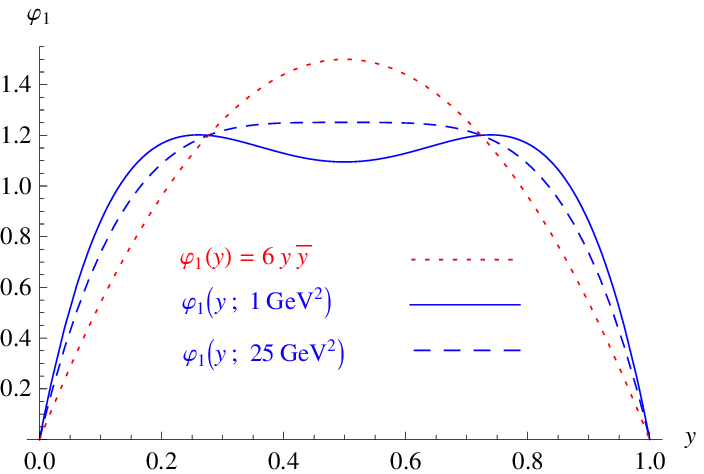}
&
\includegraphics[width=0.325\textwidth]{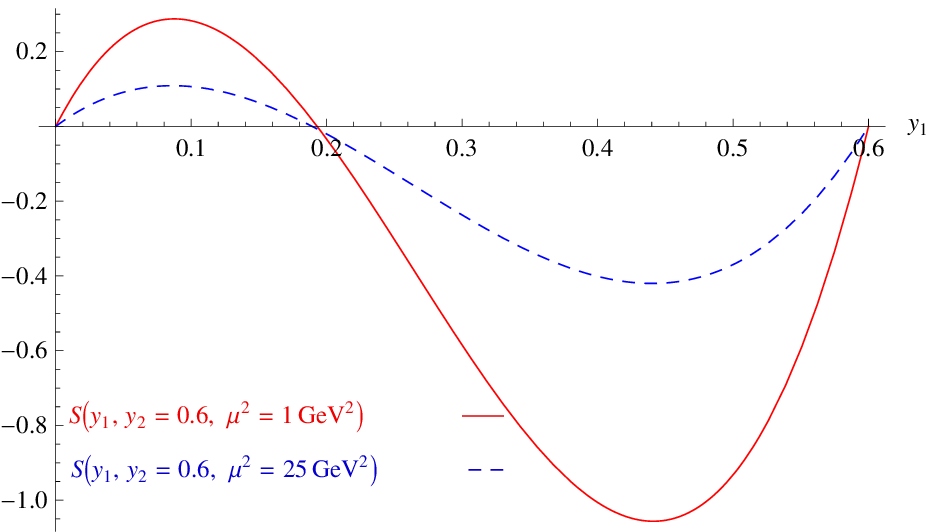}&
\includegraphics[width=0.325\textwidth]{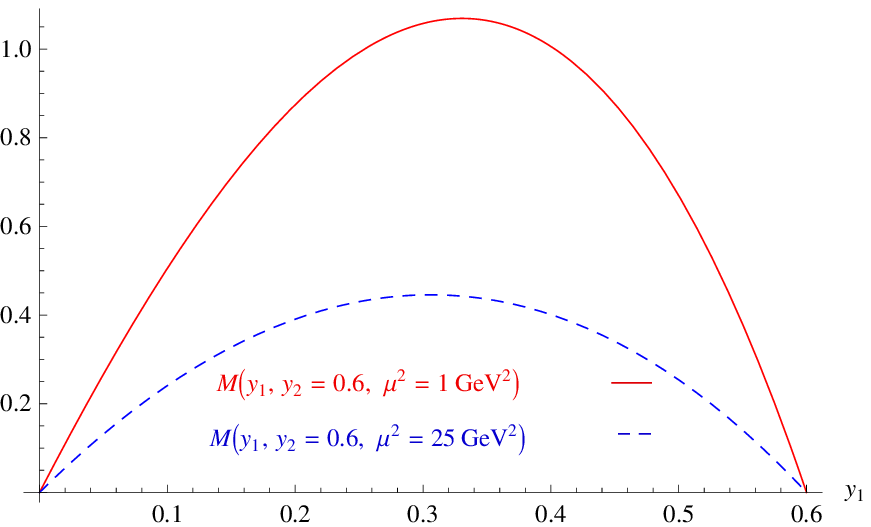}
\end{tabular}
	\caption{The three independent DAs. Left panel:  $\varphi_1(y;\mu^2)$ as a function of $y$; the red (dotted) line is for the asymptotical DA, the blue (solid) line is for  $\mu^2=1$ GeV$^2$, and the blue (dashed) line is for  $\mu^2=25$ GeV$^2$. Center panel: $S(y_1,y_2=0.6;\mu^2)$ as a function of $y_1$;  the red (solid) line is for $\mu^2=1$ GeV$^2$ and the blue (dashed) line is for $\mu^2=25$ GeV$^2$. Right panel: $M(y_1,y_2=0.6;\mu^2)$ as a function of $y_1$;  the red (solid) line is for $\mu^2=1$ GeV$^2$ and the blue (dashed) line is for $\mu^2=25$ GeV$^2$.}
\label{Fig:phi1Q_S_M}
\end{figure}

\begin{figure}[htbp]
	\begin{tabular}{cc}
	\hspace{-.3cm}	\includegraphics[width=0.45\textwidth]{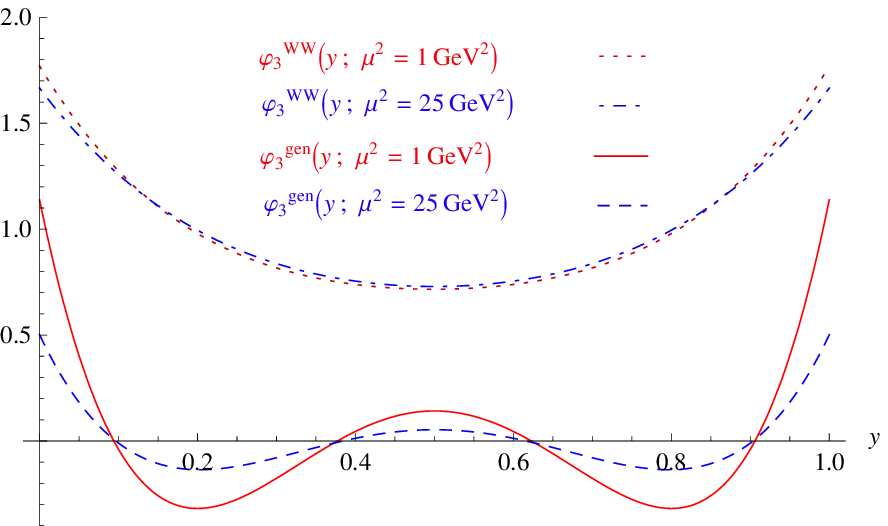}
& \hspace{.3cm}
\includegraphics[width=0.45\textwidth]{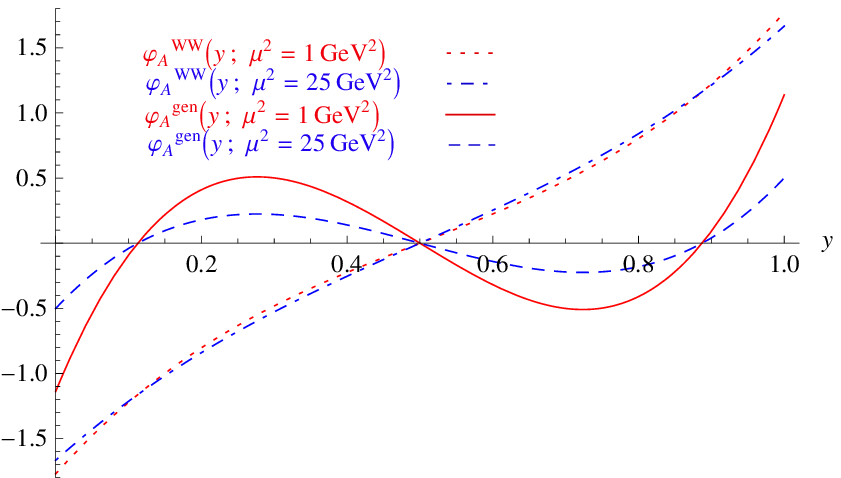}
\end{tabular}
	\caption{The two DAs  $\varphi_3$ (left panel) and $\varphi_A$ (right panel) as a function of $y$.  The red (dotted) line is the WW contribution with $\mu^2=1$ GeV$^2$, the blue (dash-dotted) line is for  the WW contribution with $\mu^2=25$ GeV$^2$, the   red (solid) line is the genuine  contribution with $\mu^2=1$ GeV$^2$, and the blue (dashed) line is the genuine  contribution with $\mu^2=25$ GeV$^2$.}
\label{Fig:phi3Q_phiAQ}
\end{figure}

\begin{figure}[htbp]
	\begin{tabular}{cc}
	\hspace{-.3cm}	\includegraphics[width=0.45\textwidth]{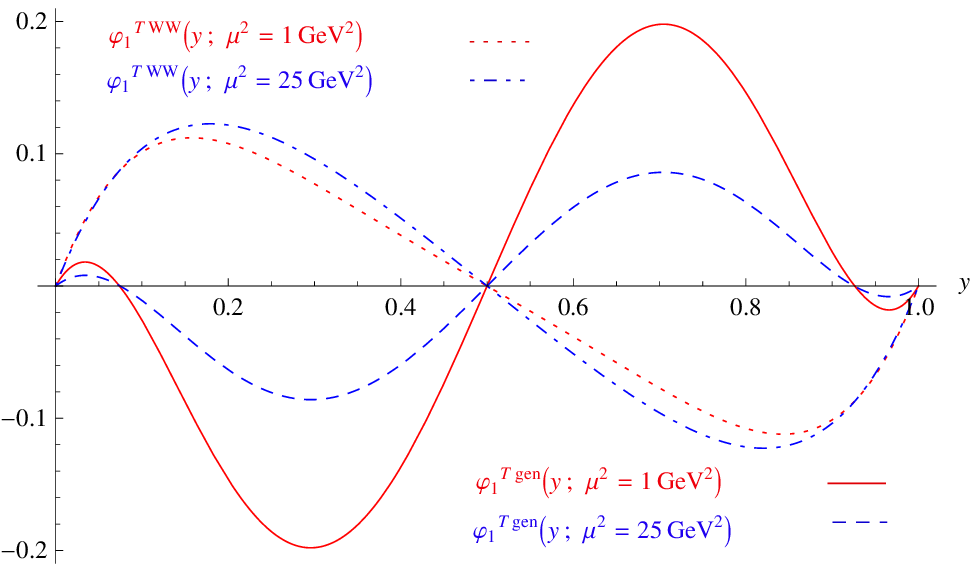}
& \hspace{.3cm}
\includegraphics[width=0.45\textwidth]{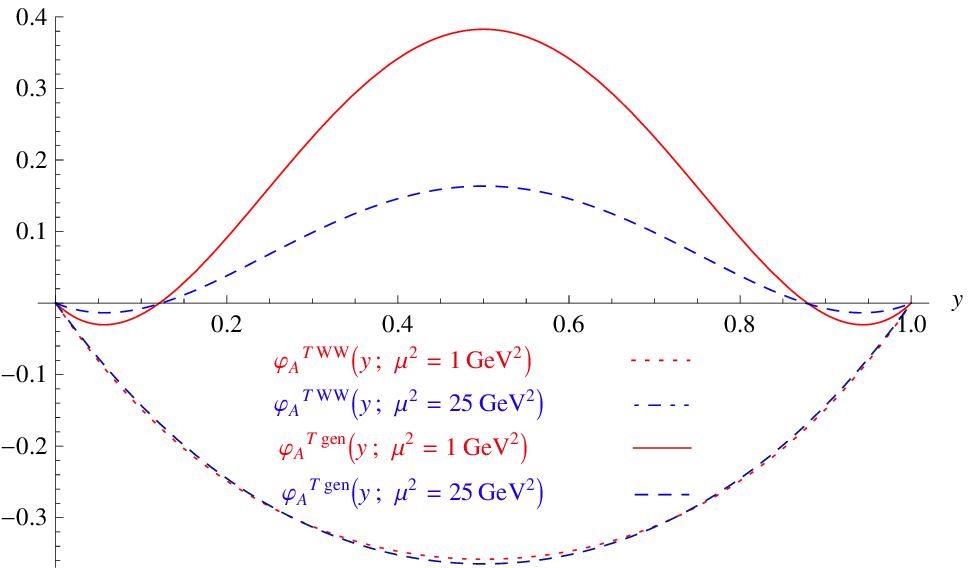}
\end{tabular}
	\caption{The two DAs  $\varphi_1^T$ (left panel) and $\varphi_A^T$ (right panel) as a function of $y$.  The red (dotted) line is the WW contribution with $\mu^2=1$ GeV$^2$, the blue (dash-dotted) line is the WW contribution with $\mu^2=25$ GeV$^2$, the red (solid) line is the genuine  contribution with $\mu^2=1$ GeV$^2$, and the blue (dashed) line is the genuine  contribution with $\mu^2=25$ GeV$^2$.}
\label{Fig:phi1TQ_phiATQ}
\end{figure}

For $a_2$, the evolution equation is
\begin{equation}
a_2(\mu^2)= a_{2}(\mu_0^2) \, L(\mu^2)^{\gamma_2/b_0}
\end{equation}
with
\begin{equation}
\label{defL}
L(\mu^2) = \frac{\alpha_s(\mu^2)}{\alpha_{s}(\mu_0^2)} = \frac{1}{1+\frac{b_0}{\pi} \alpha_{s}(\mu_0^2) \ln(\mu^2/\mu_0^2)}
\end{equation}
where $b_0=(11 N_c-2 N_f)/3\,,$
$\gamma_n=4 C_F\left (\psi(n+2)+\gamma_E-\frac{3}{4}-\frac{1}{2 (n+1)(n+2)}\right)$
and
$\psi(n) = -\gamma_E + \sum^{n+1}_{k=1} 1/k\,.$
For the $f^{A}_{3\rho}$ coupling constant, the evolution is given by
\begin{equation}
 f^{A}_{3\rho}(\mu^2)=f^{A}_{3\rho}(\mu_0^2) L(\mu^2)^{\Gamma^{-}_2/b_0}
 \end{equation}
with $\Gamma_2^- = -\frac{C_F}{3}+3 \, C_g$ ($C_g = N_c$).
The couplings
$f^{V}_{3\rho}$ and $\omega^{A}_{ \{ 0,1\} }(\mu^2) f^{A}_{3\rho}(\mu^2)$ enter a matrix evolution equation \cite{Ball:1998sk}. Defining
\beqa
\label{defV}
 V(\mu^2)=\left(
 \begin{array}{clrr}
       \omega^{V}_{[0,1]} f^{V}_{3\rho}(\mu^2) -  \omega^{A}_{ \{ 0,1\} }(\mu^2) f^{A}_{3\rho}(\mu^2)\\
\\
       \omega^{V}_{[0,1]} f^{V}_{3\rho}(\mu^2) + \omega^{A}_{ \{ 0,1\} }(\mu^2) f^{A}_{3\rho}(\mu^2)
\end{array}
\right)
\eqa
it reads
\begin{equation}
\label{V_ev}
V(\mu^2)=L(\mu^2)^{\Gamma^{+}_3/b_0} V(1)
\end{equation}
with $\Gamma^{+}_3$ given by
\beq
\label{defGamma3}
 \Gamma^{+}_3=\left(
 \begin{array}{clrr}
       \frac{8}{3}C_F+\frac{7}{3}C_g & \frac{2}{3}C_F-\frac{2}{3}C_g \\
\\
       \frac{5}{3}C_F-\frac{4}{3}C_g & \frac{1}{6}C_F+4 C_g
\end{array}
\right)\,.
\eq
Hence we get the dependence of $f^{V}_{3\rho}$ and $\omega^{A}_{ \{ 0,1\} }$ by diagonalizing the system. In Fig.~\ref{Fig:phi1Q_S_M} we display the three independent DAs $\varphi_1$ (left panel), $S$ (center panel), $M$ (right panel). Figure \ref{Fig:phi3Q_phiAQ} shows the DAs $\varphi_3$ (left panel)  and $\varphi_A$ (right panel), and Fig.~\ref{Fig:phi1TQ_phiATQ} shows the DAs $\varphi_1^T$ (left panel)  and $\varphi_A^T$ (right panel) as a function of their longitudinal variables.

%

%

%

These figures exhibit the non-negligible effect of QCD evolution on DAs, in particular, for the genuine
twist-3 contributions.

\section{Integration over $\kg$ for $T_{00}$ and $T_{01}$}
\label{subsec:T00_T01_general}

In this appendix we describe the method used to evaluate the integrals over $\kg$ in Eqs.~(\ref{defT00_general}, \ref{defT01_general}) when no infra-red cutoff is imposed.
Let ($\textbf{u}_1$,$\textbf{u}_2$) be the  orthonormal basis such as $\rg= r \:\textbf{u}_1$, and then $\textbf{k}-\textbf{r}=(k_1-r,k_2)$.
 In that case, a residue method, as described in Ref.\cite{Ginzburg:1985tp}, can be applied for the $k_1$ integration.
In this basis, the amplitudes read
\beqa
\label{T00_gen_coord}
&&\hspace{-.3cm}T_{00}=\frac{is \, C_F 2Q AB}{(2\pi)^2 (M^2+(r/2)^2)} \int^{1}_{0}\!\!dy\:y\bar{y}\,\varphi_1(y,\mu^2)\int^{\infty}_{-\infty}\!\!d k_1\int^{\infty}_{-\infty}\!\!d k_2 \frac{1}{k_1^2+k_2^2}\frac{1}{(k_1-r)^2+k_2^2}\frac{(k_1-r/2)^2+k_2^2-(r/2)^2}{(k_1-r/2)^2+k_2^2+M^2}\nonumber\\
&&\times
\left\{\!\frac{1}{(y r)^2+y\bar{y}\,Q^2}+\frac{1}{(\bar{y} r)^2+y\bar{y}\,Q^2}-\frac{1}{(k_1-y r)^2+k_2^2+y\bar{y}\,Q^2}-\frac{1}{(k_1-\bar{y} r)^2+k_2^2+y\bar{y}\,Q^2}\!\right\}
\eqa
and
\beqa
&&\hspace{-.4cm}T_{01}\!=\!\frac{is\, C_F AB}{(2\pi)^2 (M^2+(r/2)^2)}\! \int^{1}_{0}\!\!\!dy\:(y-\bar{y})\varphi_1(y,\mu^2)\!\!\int^{\infty}_{-\infty}\!\!\!\!d k_1\!\!\int^{\infty}_{-\infty}\!\!\!\!d k_2 \frac{1}{k_1^2+k_2^2}\frac{1}{(k_1-r)^2+k_2^2}\frac{(k_1-r/2)^2+k_2^2-(r/2)^2}{(k_1-r/2)^2+k_2^2+M^2} \nonumber\\
&&\times
\left\{\frac{y r \, e_1}{(y r)^2+y\bar{y}\,Q^2}-\frac{\bar{y} r \,e_1}{(\bar{y} r)^2+y\bar{y}\,Q^2}+\frac{e_1 (k_1-y r)+e_2 k_2}{(k_1-y r)^2+k_2^2+y\bar{y}\,Q^2}-\frac{e_1(k_1-\bar{y}r)+e_2 k_2}{(k_1-\bar{y} r)^2+k_2^2+y\bar{y}\,Q^2}\right\}\,,
\eqa
where $e_1$ and $e_2$ are the components of the transverse polarization of the $\gamma^{*}$ in the basis ($\textbf{u}_1$,$\textbf{u}_2$).
We define the integrands $f_{00}$ and $f_{01}$ as
\begin{equation}
\label{T00_F00}
T_{00}=\frac{is C_F 2Q AB}{(2\pi)^2 (M^2+(\textbf{r}/2)^2)} \int^{1}_{0}dy\:y\bar{y}\,\varphi_1(y; \mu^2)\int^{\infty}_{-\infty}d k_1\int^{\infty}_{-\infty}d k_2\, F_{00}(k_1,k_2,y)
\end{equation}
\begin{equation}
\label{T00_F01}
T_{01}=\frac{is \, C_F AB}{(2\pi)^2 (M^2+(r/2)^2)} \int^{1}_{0}dy\:(y-\bar{y})\,\varphi_1(y;\mu^2)\int^{\infty}_{-\infty}d k_1\int^{\infty}_{-\infty}\,d k_2 \, F_{01}(k_1,k_2,y)\,.
\end{equation}
We perform a shift of the variables: $\textbf{k}\rightarrow (\textbf{x}+\textbf{u}_1)\frac{r}{2}$ i.e $k_1 \rightarrow x_1$, $k_2 \rightarrow x_2$. The shift symmetrizes and rescales the momenta of the gluons, which are zeros for $x_1=\pm 1$ and $x_2 = 0$.
The integrands then read
\begin{multline}
\label{def_f00}
f_{00}(x_1,x_2,y)=\frac{4}{r^2}\frac{1}{(x_1+1)^2+x_2^2}\,\frac{1}{(x_1-1)^2+x_2^2}\, \frac{x_1^2+x_2^2-1}{g(M^2)+x_1^2}\\
\times \left\{\frac{1}{(y r)^2+y\bar{y}Q^2}+\frac{1}{(\bar{y} r)^2+y\bar{y}Q^2}-\frac{4}{r^2}\left[\frac{1}{(x_1+(y-\bar{y}))^2+g(y\bar{y}\, Q^2)}+\frac{1}{(x_1-(y-\bar{y}))^2+g(y\bar{y}\,Q^2)}\right]\right\}
\end{multline}
and
\begin{multline}
\label{def_f01}
f_{01}(x_1,x_2,y)=\frac{4}{r^2}\frac{1}{(x_1+1)^2+x_2^2}\frac{1}{(x_1-1)^2+x_2^2} \frac{x_1^2+x_2^2-1}{g(M^2)+x_1^2} \\
\times\left\{\frac{y \, r \, e_1}{(y r)^2+y\bar{y}Q^2}+\frac{\bar{y}\,r \, e_1}{(\bar{y} r)^2+y\bar{y}Q^2}-\frac{2}{r}\left[\frac{x_1+(y-\bar{y})}{(x_1+(y-\bar{y}))^2+g(y\bar{y}\,Q^2)}+\frac{x_1-(y-\bar{y})}{(x_1-(y-\bar{y}))^2+g(y\bar{y}\,Q^2)}\right]\right\}
\end{multline}
with
\beq
\label{defg}
g(v)=\frac{4 v+r^2 x_2^2}{r^2}
\eq
 and $f_{00}\,,$ $f_{01}$ being defined such that
\begin{equation}
\label{T00x}
T_{00}=\frac{is \,C_F 2Q AB}{(2\pi)^2 (M^2+(r/2)^2)} \int^{1}_{0}dy\:y\bar{y}\,\varphi_1(y,Q^2)\int^{\infty}_{-\infty}d x_1\int^{\infty}_{-\infty}d x_2\, f_{00}(x_1,x_2,y)
\end{equation}
and
\begin{equation}
\label{T01x}
T_{01}=\frac{is \,C_F AB}{(2\pi)^2 (M^2+(r/2)^2)} \int^{1}_{0}dy\:(y-\bar{y})\,\varphi_1(y,Q^2)\int^{\infty}_{-\infty}d x_1\int^{\infty}_{-\infty}d x_2\, f_{01}(x_1,x_2,y)\,.
\end{equation}

\subsection{Ratio $T_{01}/T_{00}$ without any infrared cutoff}
\label{subsubsec:No_IR_cutoff}

Since the integrands in (\ref{T00x}) and   (\ref{T01x}) oscillate quickly, we have used a  method where the integration over $x_1$ can be analytically performed in order to avoid numerical integration issues.
We now integrate over the variable $x_1$ using the residue method. The poles of the integrands are the same. The poles enclosed in the below contour line for $x_2 \geq 0$ are
\beqa
x_1=\pm(y-\bar{y})-i\sqrt{g(Q^2 y \bar{y})}\,, \qquad x_1=\pm1-ix_2\,, \qquad x_1=-i\sqrt{g(M^2)}\,.
\eqa
As the integrands are symmetric under $x_2 \leftrightarrow -x_2$, the result is the same for $x_2\leq0$.
The remaining integrals over $x_2$ and $y$ are then
\begin{equation}
T_{00}(r,Q,M)=\frac{is \,C_F 2 Q AB}{(2\pi)^2 (M^2+(r/2)^2)} \int^{1}_{0}dy\:y\bar{y}\varphi_1(y,Q^2)\int^{\infty}_{-\infty}d x_2 (-2i\pi) \sum^{5}_{i=1}Res_i[f_{00}]
\end{equation}
\begin{equation}
T_{01}(r,Q,M)=\frac{is \, C_F AB}{(2\pi)^2 (M^2+(r/2)^2)} \int^{1}_{0}dy\:(y-\bar{y})\varphi_1(y,Q^2)\int^{\infty}_{-\infty}d x_2 (-2i\pi) \sum^{5}_{i=1}Res_i[f_{01}]\,,
\end{equation}
where the explicit expressions for the residues $Res_i[f_{00}]$ and $Res_i[f_{01}]$ are too lengthy to be displayed here.
We then integrate numerically over $x_2\in [0,\infty]$ and $y$ and finally multiply the result due to the symmetry of the pole structure by 2.

 \subsection{Ratio $T_{01}/T_{00}$ with an infrared cutoff}
\label{subsubsec:IR_cutoff}

In order to integrate over $k_1$ and $k_2$ with an infrared cutoff $\lambda$, we will change the variables $x_1$ and $x_2$ by the distances of the point $(x_1,x_2)$ from the singularities in (-1,0) and (1,0). Let $b_1$ and $b_2$ be these distances such that
\beq
\label{def_b1_b2}
b_1^2=(x_1+1)^2+x_2^2 \qquad
b_2^2=(x_1-1)^2+x_2^2\,.
\eq
We have two solutions for $(x_1,x_2)$, one restricted to the upper half-plane,
\beq
\label{x1_sup}
x_1=\frac{1}{4} (b_1^2-b_2^2)\,, \qquad x_2=\frac{1}{4} \sqrt{(b_1+b_2+2)(b_1+b_2-2)(b_1-b_2+2)(-b_1+b_2+2)}\,,
\eq
and one restricted to the lower half-plane,
\beq
\label{x1_inf}
x_1=\frac{1}{4} (b_1^2-b_2^2)
\,, \qquad
x_2=-\frac{1}{4} \sqrt{(b_1+b_2+2)(b_1+b_2-2)(b_1-b_2+2)(-b_1+b_2+2)}\,.
\eq
We can restrain the computations to the upper half-plane because the integrands of (\ref{T01x}) and (\ref{T00x}) are invariant in $x_2 \leftrightarrow -x_2$.
The existence of both solutions requires that
\begin{equation}
(b_1+b_2+2)(b_1+b_2-2)(b_1-b_2+2)(-b_1+b_2+2)\geq 0
\end{equation}
or, equivalently, at fixed $b_1$, $b_2\leq b_1+2$ and $b_2\geq \left|b_1-2 \right|\,.$
This is the condition for the two circles centering in (-1,0) and (1,0) and of radius $b_1$ and $b_2$, respectively, to cross each other.
The Jacobian of the transformation is
\beq
\label{jacobian}
\frac{2 \, b_1 b_2}{\sqrt{(b_1+b_2+2)(b_1+b_2-2)(b_1-b_2+2)(-b_1+b_2+2)}}\geq 0
\eq
The infrared cutoff is included by a representation of the Heaviside distribution:
\newline
\begin{equation}
\theta(b_1-\beta_{cut};k)=\frac{1}{1+e^{-2k(b_1-\beta_{cut})}}\,,
\label{Heaviside}
\end{equation}
where
$\beta_{cut}$ is the cutoff for $b_1$ and $b_2$.  The link with the IR cutoff in GeV is $\lambda = \beta_{cut} \frac{r}{2}$. This ensures the stability of the numerical evaluation of the $b_1$ and $b_2$ integrations.
Then the amplitudes read
\beqa
\label{T00_lambda}
&&T_{00}(\lambda;k)=\frac{is \, C_F 2Q AB}{(2\pi)^2 (M^2+(\textbf{r}/2)^2)} \int^{1}_{0}dy\, y\bar{y}\, \varphi_1(y;\mu^2) \int^{\infty}_{0}d b_1\int^{b_1+2}_{\left|b_1-2\right|}d b_2 \, \theta(b_1-\beta_{cut}(\lambda);k) \, \theta(b_2-\beta_{cut}(\lambda);k) \nonumber\\
&& \times \frac{2 b_1 b_2}{\sqrt{(b_1+b_2+2)(b_1+b_2-2)(b_1-b_2+2)(-b_1+b_2+2)}} f_{00}(y,x_1(b_1,b_2),x_2(b_1,b_2))
\eqa
\beqa
\label{T01_lambda}
&&T_{01}(\lambda;k)=\frac{is \, C_F 2Q AB}{(2\pi)^2 (M^2+(\textbf{r}/2)^2)} \int^{1}_{0}dy\:(y-\bar{y})\varphi_1(y;\mu^2) \int^{\infty}_{0}d b_1\int^{b_1+2}_{\left|b_1-2\right|}d b_2 \theta(b_1-\beta_{cut}(\lambda);k)\\
&&\times\, \theta(b_2-\beta_{cut}(\lambda);k) \frac{2 \, b_1 b_2}{\sqrt{(b_1+b_2+2)(b_1+b_2-2)(b_1-b_2+2)(-b_1+b_2+2)}} f_{01}(y,x_1(b_1,b_2),x_2(b_1,b_2))\,.\nonumber
\eqa
The integrations are performed numerically over $b_1$, $b_2$, and $y$. The constant $k$ is chosen to be equal to 10.
The width of model (\ref{Heaviside}) of the Heaviside distribution equals $r\times\frac{1}{k}=\frac{r}{10}\,$GeV.
This ensures the stability of the computation without significantly affecting our results.

\bibliographystyle{prsty}




\end{document}